\documentclass[10pt]{iopart}
\usepackage{iopams}
\usepackage{color}
\usepackage{graphicx}
\usepackage{natbib}
\usepackage{bm}
\usepackage{ulem}

\def\vts{v_{{\rm t}s}}

\def\vti{v_{{\rm ti}}}

\def\rhoLsO{\rho_{{\rm c}0s}}

\def\rhoLi{\rho_{{\rm ci}}}

\def\omLsO{\omega_{{\rm c}0s}}

\def\omLiO{\omega_{{\rm c0i}}}

\def\omegad{\omega_{\rm d}}
\def\omegads{\omega_{{\rm d}s}}
\def\omegadi{\omega_{\rm di}}

\def\vAO{v_{{\rm A0}}}

\def\omAO{\omega_{{\rm A0}}}

\def\me{m_{\rm e}}

\def\Ts{\mathcal{T}_s}
\def\Te{\mathcal{T}_{\rm e}}
\def\Ti{\mathcal{T}_{\rm i}}

\def\tauei{\tau_{\rm ei}}

\def\feqs{f_{0s}}

\def\nsO{n_{s0}}

\def\L{\mathcal{L}}
\def\T{\mathcal{T}}
\def\O{\mathcal{O}}
\def\enr{\mathcal{E}}

\def\dphi{\delta\phi}
\def\dpsi{\delta\psi}
\def\dBp{\delta B_\parallel}

\def\eps{\varepsilon}

\def\sgnVp{\hat{\sigma}}

\def\krhoi{\hat{k}_{\rm i}}

\def\krhoiO{\hat{k}_{\rm 0i}}
\def\krhosO{\hat{k}_{{\rm 0}s}}

\def\nablab{\bm\nabla}

\definecolor{gray}{rgb}{0.5,0.5,0.5}
\definecolor{dred}{rgb}{0.5,0.0,0.0}
\definecolor{dgreen}{rgb}{0.0,0.5,0.0}
\definecolor{dblue}{rgb}{0.0,0.0,0.5}

\begin{document}

\title[$\alpha$TAE: I. Ideal MHD and FLR effects]{Pressure-gradient-induced Alfv\'{e}n eigenmodes: \\ I. Ideal MHD and finite ion Larmor radius effects}

\author{A Bierwage$^{1,\footnotemark}$, L Chen$^{1,2}$ and F Zonca$^3$}
\address{$^1$ Department of Physics and Astronomy, University of California, \\ Irvine, CA 92697, USA}
\address{$^2$ Institute for Fusion Theory and Simulation, Zhejiang University, \\ Hangzhou, People's Republic of China}
\address{$^3$ Associazione EURATOM-ENEA sulla Fusione, CP 65-00044 Frascati, \\ Rome, Italy}
\eads{\mailto{andreas@bierwage.de} \mailto{liuchen@uci.edu} \mailto{zonca@frascati.enea.it}}

\footnotetext{Present address: Associazione EURATOM-ENEA sulla Fusione, CP 65-00044 Frascati, Rome, Italy.}


\begin{abstract}
In the second magnetohydrodynamic (MHD) ballooning stable domain of a high-beta tokamak plasma, the Schr\"{o}dinger equation for ideal MHD shear Alfv\'{e}n waves has discrete solutions corresponding to standing waves trapped between pressure-gradient-induced potential wells. Our goal is to understand how these so-called $\alpha$-induced toroidal Alfv\'{e}n eigenmodes ($\alpha$TAE) are modified by the effects of finite Larmor radii (FLR) and kinetic compression of thermal ions in the limit of massless electrons. In the present paper, we neglect kinetic compression in order to isolate and examine in detail the effect of FLR terms. After a review of the physics of ideal MHD $\alpha$TAE, the effect of FLR on the Schr\"{o}dinger potential, eigenfunctions and eigenvalues are described with the use of parameter scans. The results are used in a companion paper to identify instabilities driven by wave-particle resonances in the second stable domain.
\end{abstract}


\section{Introduction}
\label{sec:intro}

The second magnetohydrodynamic (MHD) ballooning stable domain \cite{Coppi80} is an attractive parameter regime which may be utilized to achieve high plasma pressures in toroidal magnetic confinement devices such as tokamaks with the goal to create thermonuclear fusion conditions. Experimental access to the second stable domain remains a challenging task requiring reliable profile control (to circumvent the ballooning unstable domain via negative magnetic shear and the average magnetic well), optimized shaping of the plasma cross-section (to maximize the average magnetic well), and utilization of potentially stabilizing mechanisms (e.g., sheared flows \cite{Taylor99}).

In the meantime, theoretical and numerical studies have advanced to explore the properties of the second stable domain. The present paper and its companion \cite{Bierwage10b} are part of this effort and were motivated by a discovery made by Hu \& Chen \cite{Hu04}: using the $s$-$\alpha$ model equilibrium \cite{Connor78}, where $s$ is the magnetic shear and $\alpha$ the normalized pressure gradient, Hu \& Chen have shown that the second ballooning stable domain is populated by discrete ideal MHD Alfv\'{e}n eigenmodes. These so-called $\alpha$-induced toroidal Alfv\'{e}n eigenmodes ($\alpha$TAE) are standing waves trapped between potential barriers. These barriers represent the modulation of magnetic curvature and shear due to the effect of a steep pressure gradient (corresponding to large $\alpha$) and cause reflections of shear Alfv\'{e}n waves (SAW).

Hu \& Chen \cite{Hu05} have further demonstrated that $\alpha$TAEs exist for negative magnetic shear and that these modes may be resonantly excited through interactions with a sparse energetic ion population. These results are based on a hybrid model where an ideal MHD bulk plasma interacts with gyrokinetic energetic ions. Our goal is to extend this earlier study by including non-ideal effects of a collisionless bulk plasma, such as finite ion Larmor radii (FLR), finite parallel electric field ($\delta E_\parallel$), and kinetic ion compression ($\delta G_{\rm i}$).

In the present paper, we exclude kinetic compression and construct a fluid limit, which we call ``FLR MHD.'' This model is not self-consistent, but it is useful to isolate the effect of FLR terms on ideal MHD $\alpha$TAEs. The questions dealt with in the present paper may be summarized as follows:
\begin{enumerate}
\item  What are $\alpha$TAEs? (a review with detailed discussion of bound state formation in ideal MHD toroidal plasma)
\item  How are $\alpha$TAE affected by FLR corrections? (analysis of the Schr\"{o}dinger potential, mode structures and eigenvalues; in particular, frequency shifts)
\item  How is the propagating component of the mode structure affected by FLR corrections? (peculiar features of shear Alfv\'{e}n continuum waves in FLR MHD)
\end{enumerate}

\noindent The results of item (ii) will enable us to identify the branches of ion-temperature-gradient (ITG)-driven Alfv\'{e}nic instabilities observed in gyrokinetic simulations solving the self-consistent model equations. This is done in the companion paper \cite{Bierwage10b}.

The FLR MHD model and the numerical methods used are described in section \ref{sec:model}. In section \ref{sec:ideal}, we review the physical origin and basic features of ideal MHD $\alpha$TAEs. In section \ref{sec:struc}, we show how the structure of the effective Schr\"{o}dinger potential in the FLR MHD model varies with the mode frequency and demonstrate that, in cases of interest, the dominant bound state component of an $\alpha$TAE eigenfunction remains essentially unchanged by FLR corrections. The eigenfrequencies of two $\alpha$TAE branches are inspected in section \ref{sec:scan}, where we analyze their dependence on $\alpha$ (pressure gradient), $k_\vartheta \rhoLi$ (poloidal wavenumber $\times$ ion Larmor radius), and $\eta_{\rm i}$ (ratio of density and temperature gradient scale lengths). It is also shown that the magnitude of the diamagnetic frequency shift depends on the mode structure. In section~\ref{sec:conclude}, the results are summarized and conclusions are drawn. The Appendix contains a detailed discussion of how (propagating) continuum waves are modified by FLR effects.

With our choice of terminology, we propose to use the term ``toroidal Alfv\'{e}n eigenmode'' (TAE) in a more general sense than is usually done. Our concept of TAE encompasses all ideal MHD eigenmodes formed inside the quasi-periodic structure of the effective Schr\"{o}dinger potential along a flux tube; which cannot be found in the cylindrical limit, only in \textit{toroidal geometry}. This incorporates effects due to finite aspect ratio ($\eps = a/R_0$), shaping of flux surfaces (ellipticity, triangularity, etc.), and distortions caused by the pressure gradient ($\alpha$). The ``classical'' TAE  is associated with the $\eps$-dependence of the magnetic field strength, $B(\theta) \approx B_0/(1 + \eps\cos\theta)$. When the $\alpha$-induced potential barriers dominate (as in the present work), we speak of $\alpha$TAEs. Naturally, there are parameter regimes where no clear distinction can be made between the ``classical'' TAE and $\alpha$TAE: typically, when $\eps$ and $\alpha$ have similar values. More generally, one can refer to Alfv\'{e}n eigenmodes (AE) as discrete bound states due to equilibrium non-uniformities causing poloidal symmetry breaking \cite{Chen07}.

\section{Physical model and numerical method}
\label{sec:model}

The present work is part of an effort to study and understand the dynamics of drift Alfv\'{e}n ballooning modes in a wide range of frequencies and wavelengths and their kinetic excitation in a high-$\beta$ plasma with toroidal geometry. The full model is given by the one-dimensional linear gyrokinetic equations based on the derivation by Chen \& Hasegawa \cite{Chen91}. There, the electromagnetic field perturbations are described by gyrokinetic Maxwell equations in terms of the magnetic flux function $\delta\psi$, the electrostatic potential $\delta\phi$, and the magnetic compression $\delta B_\parallel$. The linear gyrokinetic Vlasov equation governs the evolution of the fluctuating part, $\delta f_s$, of the total distribution function, $f_s = \feqs + \delta f_s$, where the \textit{subscript} $s$ labels the particle species (not to be confused with the magnetic shear).

Here, we utilize a reduced model which is obtained from \cite{Chen91} (and more recent formulations in \cite{Zonca06b, Bierwage08}) by excluding energetic ions, ignoring kinetic thermal ion compression, including magnetic compression only at lowest order, and assuming a Maxwellian equilibrium distribution, $\feqs$. We also ignore the variation of the magnetic field strength along a flux tube by letting $B = B_0$, so there are no magnetically trapped particles and no toroidicity-induced gap in the Alfv\'{e}n continuum. However, magnetic curvature and $\nablab B$ drifts are properly accounted for. The reduced equations are presented in section \ref{sec:model_incomp}, where the connection with the original gyrokinetic model is shown and the physical meaning of individual terms is given. In section \ref{sec:model_S}, the three equations are combined into one single SAW equation, written in the form of a linear time-independent Schr\"{o}dinger equation. This equation is solved with a standard shooting method, the appropriate boundary condition for which is given in section \ref{sec:model_bc}.

\subsection{FLR MHD equations for shear Alfv\'{e}n waves}
\label{sec:model_incomp}

With the use of the ballooning formalism \cite{Connor78, Coppi77, Lee77, Pegoraro78, Dewar81}, the poloidal angle coordinate, $\vartheta\in[0,2\pi]$, is mapped onto an infinite covering space, $\theta \in(-\infty,\infty)$. The coordinate $\theta$ effectively measures the distance along the field line normalized by $q R_0$, with $R_0$ being the major radius of the magnetic axis and $q$ the safety factor measuring the average field line pitch. Thus, the connection length in $\theta$ is $2\pi$. For convenience, we write the equations in Laplace-transformed form ($\partial_t \rightarrow -i\omega$, with $\omega = \omega_{\rm r} + i\gamma$). We use SI units and define the following coefficients and parameters:
\begin{eqnarray}
\fl & Q_s = \frac{\omega_{*s}^T - \omega}{\T_s}, \quad 
\omega_{*s}^T = \omega_{*s}\left[1 + \eta_s\left(\frac{\enr}{\T_s} - \frac{3}{2}\right)\right], \quad
\omega_{*ps} = \omega_{*s}(1 + \eta_s), &
\label{eq:defs}
\\
\fl & \omega_{*s} = \frac{k_\vartheta\T_s}{\omLsO L_n}, \quad
\eta_s = \frac{\T_s'/\T_s}{\nsO'/\nsO}, \quad
L_n^{-1} = -\frac{\nsO'}{\nsO}, \quad
\eps_n = \frac{L_n}{R_0}, \quad
\tau_{{\rm e}s}^T = \frac{\me\Te}{m_s\T_s}, & \nonumber
\\
\fl & \omegads = \frac{\Omega_\kappa}{\omLsO} \left(v_\parallel^2 + \mu B_0\right) + \frac{\Omega_p}{\omLsO} \mu B_0, \quad
\Omega_\kappa = \frac{k_\vartheta}{R_0} g, \quad
\Omega_p = -\frac{k_\vartheta \alpha}{2 q^2 R_0}, \quad
\omLsO = \frac{e_s B_0}{m_s}, & \nonumber
\\
\fl & k_\perp = \sqrt{f} k_\vartheta, \quad
\krhosO = k_\vartheta \vts/\omLsO, \quad
\lambda_s = k_\perp \rhoLsO = \sqrt{f} \krhoiO v_\perp / \vts, \quad
b_{0s} = f \krhosO^2, & \nonumber
\\
\fl & f = 1 + h^2, \quad
g = \cos\theta + h\sin\theta, \quad
h = s(\theta-\theta_k) - \alpha\sin\theta. & \nonumber
\end{eqnarray}

\noindent Here, $\nsO(r)$ is the unperturbed number density, $\T_s(r) = \vts^2(r)$ corresponds to 2/3 of the thermal energy per unit mass, $\enr = (v_\perp^2 + v_\parallel^2)/2$ is the kinetic energy and $\mu = v_\perp^2/(2B)$ the magnetic moment. $\omega_{*s}$ is the diamagnetic frequency associated with the density gradient, $\omegads$ the magnetic drift frequency, and $\omLsO$ the cyclotron frequency at the magnetic axis. The quantities $f$, $g$ and $h$ describe the flux tube geometry in the shifted-circle model equilibrium in terms of the parameters $\alpha = -q^2 R_0 \beta'$ and $s = rq'/q$, where $\beta = 2\mu_0 P/B_0^2$ is the ratio between thermal and magnetic pressure, and the prime denotes a radial derivative ${\rm d}/{\rm d}r$. Apart from a brief review of the effect of nonzero $\theta_k$ (variable radial envelope) in section~\ref{sec:ideal}, we consider the case $\theta_k = 0$. Here and in the following, the attributes parallel ($\parallel$) and perpendicular ($\perp$) refer to the direction relative to the equilibrium magnetic field, the toroidal component of which is assumed to be dominant as in typical tokamaks.

Following standard procedures, the adiabatic and convective responses are separated through the substitution
\begin{equation}
\delta f_s = -\frac{e_s\feqs}{m_s\Ts}\left( \dphi + \frac{Q_s\Ts}{\omega} J_0\dpsi e^{iL_k} \right) + \frac{\delta G_s}{\omega} e^{iL_k};
\label{eq:model_vlasov_gke1}
\end{equation}

\noindent where terms involving $\partial_\mu\feqs$ are omitted since the equilibrium distribution, $\feqs = \nsO (2\pi\Ts)^{-3/2} \exp(-\enr/\T_s)$, is isotropic. $J_n(\lambda_s)$ is the Bessel function of order $n$ introduced by the gyroaverage. $L_k = -{\bm k}_\perp\cdot(\hat{\bm b}\times{\bm v}_\perp)/\omLsO$ is the generator of the coordinate transformation between guiding center and particle variables, and $e^{iL_k}$ turns into another $J_0$ when gyroaveraged. The quantity $\delta G_s$ captures the compressional part of the non-adiabatic component of the particle response (in short, kinetic compression). Electrons are approximated as a massless fluid, so that $\delta G_{\rm e} = 0$. The evolution of kinetic ion compression, $\delta G_{\rm i}$, is governed by the gyrokinetic equation \cite{Chen91}.

In the present study, we wish to ignore kinetic compression, so we let $\delta G_{\rm i}=0$. This yields the following fluid equation, which we refer to as FLR MHD model:
\begin{eqnarray}
\fl 0 =& \underbrace{\frac{k_\vartheta^2}{(q R_0)^2} \frac{\partial}{\partial\theta}\left(f \frac{\partial\delta\psi}{\partial\theta}\right)}\limits_{\rm FLB}
\underbrace{- \mu_0 \left<\frac{e^2}{m}(1 - J_0^2) Q f_0 \right>_{\rm i} \omega\delta\phi}\limits_{\rm inertia\; (ideal\; MHD\; +\; FLR)}
\underbrace{- \Omega_p\left[(\Omega_p + 2\Omega_\kappa)\delta\psi + \omega\delta B_\parallel\right]}\limits_{\rm MPC\; +\; MFC\; (drift{\tt -}kinetic)}
\label{eq:model_maxw_vort1}
\\
\fl & \underbrace{- \mu_0\left<\frac{e^2}{m}\omegad(1 - J_0^2) Q f_0 \right>_{\rm i} \delta\psi
- \frac{\mu_0}{B}\left<e \mu B \left( 1 - \frac{2J_1}{\lambda} J_0\right) Q f_0 \right>_{\rm i} \omega\delta B_\parallel}\limits_{\rm MPC\; +\; MFC\; (FLR\; correction)}, \nonumber
\\
\fl 0 =& \left<\frac{e^2}{m} \partial_{\mathcal E} f_0\right>_{\rm i} \omega(\delta\phi - \delta\psi) 
+ \left<\frac{e^2}{m} (1 - J_0^2) Q f_0\right>_{\rm i} \delta\psi,
\label{eq:model_maxw_qn1}
\\
\fl 0 =& \omega\delta B_\parallel + \Omega_p\delta\psi;
\label{eq:model_maxw_bp1}
\end{eqnarray}

\noindent where $\left<...\right> = \int{\rm d}^3 v = \sum_{\sgnVp}\int{\rm d}\enr\int{\rm d}\mu B/|v_\parallel|$ is the velocity space integral. Equation~(\ref{eq:model_maxw_vort1}) is the so-called vorticity equation, which is obtained through the combination of the parallel Amp\`{e}re's law with the continuity equation \cite{Chen91, Zonca06b}. Its individual terms describe field line bending (FLB), inertia (with FLR), MHD particle compression (MPC), and magnetic field compression (MFC). Both MPC and MFC are static compression effects associated with toroidal curvature and finite $\beta$. Equation~(\ref{eq:model_maxw_qn1}) is the quasi-neutrality condition, which originally read $\sum_s\left<e_s\delta f_s\right> = 0$. Equation (\ref{eq:model_maxw_bp1}) is the perpendicular Amp\`{e}re's law reduced to an equation for the total (magnetic plus thermal) pressure balance for convective displacements. Note that $\omega\dBp = -\Omega_p \dpsi$ effectively eliminates $\Omega_p$ (high-$\beta$ correction to the curvature drift) from $\omegadi$ at the order considered and reduces the drift-kinetic part of the ``MPC+MFC'' term to $-2\Omega_p\Omega_\kappa\delta\psi = \alpha g k_\vartheta^2/(q R_0)^2\delta\psi$ (``ideal MHD ballooning term'').

Strictly speaking, the kinetic compression terms neglected here have a non-zero MHD limit if one goes to a regime where $|\omega| \gg |\omegadi|$ and $|\omega| \gg |k_\parallel v_\parallel|$. However, in the parameter regime we are dealing with, these conditions are only satisfied for the low-energy part of the ion distribution. Thus, one cannot integrate over the entire velocity space as is usually done to compute the fluid limit (e.g., when estimating the width of the kinetic thermal ion gap \cite{Zonca96}); otherwise, the estimate will be wrong and one may as well neglect it altogether, as we choose to do here. The FLR MHD model may consequently be regarded as ``incompressible FLR MHD,'' in the sense that the ion sound branch, compressibility associated with finite aspect ratio (trapped particles), and wave-particle interactions are not included.

\subsection{FLR MHD equation in Schr\"{o}dinger form}
\label{sec:model_S}

In order to write the equations in dimensionless form, we employ the following normalizations:
\begin{eqnarray}
\hat{\omega} = \frac{\omega}{\omAO}, \quad
\hat{\omega}_{*{\rm i}} = \frac{\omega_{*{\rm i}}}{\omAO} = \frac{q \krhoiO \hat{v}_{\rm ti}}{\eps_n}, \quad
\hat{\Omega}_\kappa = \frac{q R_0 \Omega_\kappa \Ti}{\omLiO\vAO} = q \krhoiO \hat{v}_{\rm ti} g(\theta),
\label{eq:norm_omgf_omgk}
\\
\hat{\omega}_{\rm ti}^2 = \hat{v}_{\rm ti}^2 = \frac{\vti^2}{\vAO^2} = \frac{\beta_{\rm i}}{2} = \frac{\alpha \eps_n}{2 q^2 [1 + \eta_{\rm i} + \tauei^T(1 + \eta_{\rm e})]}; \nonumber
\end{eqnarray}

\noindent where $\omega_{\rm A0} = \vAO/(qR_0)$ is the Alfv\'{e}n frequency. In the following, the hats will be neglected on all quantities except $\krhoiO$.

The vorticity equation (\ref{eq:model_maxw_vort1}) is combined with equations (\ref{eq:model_maxw_qn1}) and (\ref{eq:model_maxw_bp1}), and brought into Schr\"{o}dinger-form through the substitution $\delta\Psi_{\rm s} = \sqrt{f} \delta\psi$, which yields
\begin{equation}
\delta\Psi_{\rm s}'' - V_{\rm eff}(\theta) \delta\Psi_{\rm s} = 0.
\label{eq:model_saw_eq}
\end{equation}

\noindent Here, $\delta\Psi_{\rm s}''$ stands for ${\rm d}^2(\delta\Psi_{\rm s})/{\rm d}\theta^2$. The effective Schr\"{o}dinger potential,
\begin{equation}
V_{\rm eff} = V + V_{{\rm m},\omega} + V_{{\rm m},\tau} + V_{\kappa,{\rm FLR}},
\label{eq:model_veff}
\end{equation}

\noindent consists of the components
\begin{eqnarray}
V &= (s-\alpha\cos\theta)^2/f^2 - (\alpha\cos\theta)/f,
\label{eq:model_saw_v}
\\
V_{{\rm m},\omega} &= \omega\left[\omega\left(1 - \frac{1 - \Gamma_0}{b_{\rm 0i}}\right)
+ \omega_{*{\rm i}}\frac{1 - \Gamma_0\Upsilon_1}{b_{\rm 0i}}\right] - \omega^2,
\label{eq:model_saw_vflr_w}
\\
V_{{\rm m},\tau} &= \frac{\tauei^T b_{\rm 0i}}{1 + \tauei^T} \left(\omega\frac{1 - \Gamma_0}{b_{\rm 0i}} - \omega_{*{\rm i}}\frac{1 - \Gamma_0\Upsilon_1}{b_{\rm 0i}}\right)^2,
\label{eq:model_saw_vflr_ep}
\\
V_{\kappa,{\rm FLR}} &= - 2 \Omega_\kappa \left[ \omega\frac{1 - \Gamma_0\Delta_1}{b_{\rm 0i}}
- \frac{\omega_{*p{\rm i}}}{b_{\rm 0i}} \left(1 - \frac{\Gamma_0\Upsilon_{2\kappa}}{1 + \eta_{\rm i}}\right)\right].
\label{eq:model_saw_vflr_wk}
\end{eqnarray}

\noindent The functions $\Gamma_0$, $\Delta_1$, $\Upsilon_1$ and $\Upsilon_{2\kappa}$ are defined as (cf.~equation (2.22) of \cite{Bierwage08})
\begin{eqnarray}
\fl \Gamma_0 = e^{-b_{\rm i}} I_0, \quad
\Delta_1 = 1 + (I_1/I_0 - 1)b_{\rm i}/2, \nonumber
\\
\fl \Upsilon_1 = 1 + (I_1/I_0 - 1) \eta_{\rm i} b_{\rm i}, \quad
\Upsilon_{2\kappa} = \Delta_1 + \left[1 + (3 I_1/I_0 - 4)b_{\rm i}/2 - (I_1/I_0 - 1)b_{\rm i}^2\right] \eta_{\rm i};
\label{eq:def_moments}
\end{eqnarray}

\noindent with $I_k(b_{\rm i})$ being the modified Bessel function $k$-th order.

The terms $\delta\Psi_{\rm s}'' - V\delta\Psi_{\rm s}$ are obtained by combining the ``FLB'' term with the drift-kinetic part of the ``MPC+MFC'' term (ideal MHD ballooning term) from equation (\ref{eq:model_maxw_vort1}). The ideal MHD potential $V$ is determined by the equilibrium field geometry parametrized by $s$ and $\alpha$. The ``inertia'' term of equation (\ref{eq:model_maxw_vort1}) is now $-(V_{{\rm m},\omega} + V_{{\rm m},\tau})$, where $V_{{\rm m},\omega}$ consists of ideal MHD inertia, $\omega^2$, plus FLR corrections, and $V_{{\rm m},\tau}$ captures the effect of non-zero parallel electric field, $\delta E_\parallel = -\partial_\theta(\dphi - \dpsi)$. The subscript ``m'' stands for mass (ion inertia). The FLR correction of the ``MPC+MFC'' terms is captured by $V_{\kappa,{\rm FLR}}$.

\subsection{Outgoing boundary condition}
\label{sec:model_bc}

At the outer boundary $\theta = \theta_{\rm max}$ of the shooting range (in the asymmetric case, $\theta_k \neq 0$, on both sides of the domain, $\theta = \pm\theta_{\rm max}$), we apply an outgoing boundary condition by matching the numerical solution to an analytical solution. The latter is obtained by solving equation (\ref{eq:model_saw_eq}) in the limit $|\krhoiO s \theta| \gg 1$, where it asymptotically approaches
\begin{equation}
0 = \delta\Psi_{\rm s}'' + \frac{X(\omega)}{(\krhoiO s \theta)^2} \delta\Psi_{\rm s} + Y(\omega) \frac{\sin\theta}{\krhoiO s \theta} \delta\Psi_{\rm s}, \qquad \left(|\theta| \gg 1/|\krhoiO s|\right).
\label{eq:saw_eq_flr_bc}
\end{equation}

\noindent The coefficients
\begin{equation}
\fl X(\omega) = (\omega + \tauei^T\omega_{*{\rm i}})(\omega - \omega_{*{\rm i}}) / (1 + \tauei^T), \quad
Y(\omega) = (\omega - \omega_{*p{\rm i}}) 2 q \vti
\label{eq:saw_eq_flr_bc_coeff}
\end{equation}

\noindent originate from $-(V_{{\rm m},\omega} + V_{{\rm m},\tau})$ and $-V_{\kappa,{\rm FLR}}$, respectively. As is shown in the Appendix, the solution of equation (\ref{eq:saw_eq_flr_bc}) which satisfies the causality constraint for an outgoing group velocity, $v_{\rm g}/\theta > 0$, is
\begin{equation}
\fl \delta\Psi_{\rm s} = \Psi_0 |\krhoiO s \theta|^{\frac{1}{2} + i\frac{\sigma C}{2}} \left(1 + \frac{Y\sin\theta}{\krhoiO s \theta}\right), \quad
\frac{\delta\Psi_{\rm s}'}{\delta\Psi_{\rm s}} \approx \frac{1}{2\theta} \left[i\sigma C + \left(1 + \frac{2Y\cos\theta}{\krhoiO s} \right)\right];
\label{eq:match_flr}
\end{equation}

\noindent where $\sigma = {\rm Re}\{C\}/|{\rm Re}\{C\}|$, $C = \sqrt{4X + 2Y^2 - 1}$ and $\Psi_0$ is a constant. Formally, we write $\delta\Psi_{\rm s}'/\delta\Psi_{\rm s} = ik_\parallel$. By matching the analytic solution given by equation (\ref{eq:match_flr}) to the numerical solution for equation (\ref{eq:model_saw_eq}), one ensures that the waves which travel through the entire simulation domain, are not reflected at the artificial boundaries.

Other than this, equation (\ref{eq:saw_eq_flr_bc}) and its solution has no physical meaning, because it only describes the balance between magnetic induction and the ion polarization current. The latter vanishes in the limit $|\krhoiO s \theta| \gg 1$, where the radial width of the perturbation becomes much smaller than the ion Larmor radius. Physically, magnetic induction will then be balanced by electron inertia or the wave is dissipated by collisions. These effects are not included in equation (\ref{eq:model_saw_eq}) and the underlying gyrokinetic Maxwell-Vlasov equations, and they are not important for the study of $\alpha$TAEs which are localized near $|\theta| \sim \O(1)$. $\alpha$TAEs may lose energy to the outgoing waves, which represents the physical effect of continuum damping, but are not otherwise affected by their presence. The physics of large $|\theta|$ become important only near marginal ideal MHD ballooning stability, where the mode structure becomes broad.

\subsection{Parameter values}
\label{sec:model_parms}

In the present paper, we analyze two cases: one with lower magnetic shear ($s=0.4$) and one with higher magnetic shear ($s=1.0$). The default parameters are listed in table \ref{tab:parms}. They were adopted (with minor changes) from two earlier linear gyrokinetic studies of Alfv\'{e}nic ITG instabilities \cite{Hirose94, Dong04}.

Note that is is not meaningful to simply compare the values of parameters $(s,\alpha)$ of the shifted-circle equilibrium model used here with parameter values in experimental configurations with non-circular flux surfaces. The location of the stability boundaries of the ideally unstable domain and the distribution of the $\alpha$TAE bands in the $s$-$\alpha$ plane are sensitive to the flux surface geometry, so it is likely that a given region in parameter space of the simple model may correspond to another region of the parameter space in a more realistic geometry. A meaningful comparison would need to consider the shape of the Schr\"{o}dinger potential at different points in parameter space, rather than the values of parameters like $s$ and $\alpha$. This may be the subject of a future study.

\begin{table}
\caption{Default physical parameters in the two cases considered in this paper: one with lower magnetic shear ($s=0.4$) and one with higher shear ($s=1.0$).}
\label{tab:parms}
\begin{indented}
\item[]
\begin{tabular}{@{}ccccccccl}
\br
$s$ & $q$ & $\eps_n$ & $\krhoiO$ & $\eta_{\rm i}$ & $\eta_{\rm e}$ & $\tauei^T$ & $\beta_{\rm i} = \beta_{\rm e}$ & adapted from \\
\mr
$0.4$ & $1.2$ & $0.175$ & $0.2$   & $2$   & $2$   & $1$ & $\approx \alpha/50$ & Hirose~\textit{et al.}~\protect\cite{Hirose94} \\
$1.0$ & $1.5$ & $0.2$   & $0.212$ & $2.5$ & $2.5$ & $1$ & $\approx \alpha/80$ & Dong~\textit{et al.}~\protect\cite{Dong04} \\
\br
\end{tabular}
\end{indented}
\end{table}

\section{Review of ideal MHD Alfv\'{e}n eigenmodes}
\label{sec:ideal}

Many qualitative properties of ideal MHD $\alpha$TAEs are preserved in FLR MHD. Thus, it is instructive to review the physics underlying $\alpha$TAEs in the simpler ideal MHD limit, where the SAW Schr\"{o}dinger equation~(\ref{eq:model_saw_eq}) reduces to
\begin{equation}
0 = \delta\Psi_{\rm s}'' - V\delta\Psi_{\rm s} + \omega^2\delta\Psi_{\rm s}
= \left[(f\delta\psi')' + \alpha g\delta\psi + \omega^2 f \delta\psi\right]/\sqrt{f}.
\label{eq:saw_ideal}
\end{equation}

\noindent In section \ref{sec:ideal_geom}, we elucidate the origin of the $\alpha$-induced potential barriers. The types of discrete shear Alfv\'{e}n eigenmodes are introduced in section \ref{sec:ideal_branches}: purely growing ideal MHD ballooning modes and damped oscillatory solutions, now called $\alpha$TAEs. The latter are described in detail in section \ref{sec:ideal_atae}. Section \ref{sec:ideal_radial} contains remarks regarding the effect of the radial envelope and global trapping.

\subsection{Equilibrium magnetic field geometry}
\label{sec:ideal_geom}

We employ the $s$-$\alpha$ model equilibrium for simplicity and comparability with earlier works (the results of which we explain in the companion paper \cite{Bierwage10b}). The model is not rigorously valid for $\alpha \gtrsim 1$, but even outside its regime of validity it captures essential qualitative properties of more accurate models \cite{Coppi79, Miller98}. The most prominent feature predicted by the $s$-$\alpha$ model is the second MHD ballooning stable domain.

The model is useful because it captures two essential ingredients: toroidicity and magnetic shear. The resulting inhomogeneities and asymmetries give rise to discrete shear Alfv\'{e}n eigenmodes which do not exist in the cylindrical limit. Writing the vorticity equation in Schr\"{o}dinger form, such as equation (\ref{eq:saw_ideal}), offers an illuminating way to express these physics in terms of familiar quantum mechanics concepts: toroidicity and magnetic shear introduce quasi-periodic barriers and wells in the effective Schr\"{o}dinger potential along a flux tube. As a consequence, the spectrum of ideal MHD shear Alfv\'{e}n waves in toroidal geometry consists of continua (waves propagating along a flux tube) and discrete eigenmodes (standing waves satisfying quantization conditions). Further geometric complications introduce additional gaps in the continuum and corresponding branches of discrete modes at various frequencies.

Speaking in simple terms, in the $s$-$\alpha$ model, the periodicity of the barriers is controlled by the magnetic shear $s$ and their height by the pressure gradient parameter $\alpha$. For the discrete eigenmodes studied here, the dominant effect of toroidicity is the one that enters via $\alpha$. We look at a regime where the $\alpha$-induced potential barriers are significantly larger than those induced by the inverse aspect ratio $a/R_0$, so we neglect the variation of $B$ and $R$ along a flux tube (along with all the effects of non-circular flux surfaces) and let $B(\theta) = B_0$ and $R(\theta) = R_0$ (constant).

The potential structure may be traced back to its physical origin as follows. The function $g = \cos\theta + h\sin\theta$ appearing in the ideal MHD ballooning term represents the sum of normal and geodesic curvatures. The factor $h$ reflects the fact that the modulation of geodesic curvature is intrinsically coupled to the local magnetic shear, which is measured by the quantity $h' = s - \alpha\cos\theta$. Noting that the term $-\alpha\cos\theta$ in $h'$ has the form of the Pfirsch-Schl\"{u}ter current \cite{NishikawaWakatani}, it is clear that the $s$-$\alpha$ model describes the modulation of local magnetic shear and geodesic curvature which is produced by currents that balance the charge separation induced by toroidal curvature drift.

Taking into account these distortions of the flux tube geometry, the Schr\"{o}dinger potential $V(\theta|s,\alpha)$ measures how much energy is needed to bend a field line at a given location $\theta$. From the functional form of $V$ defined in equation (\ref{eq:model_saw_v}), the numerator of which is $(s - \alpha\cos\theta)^2 - f\alpha\cos\theta$ (with $f \geq 1$), it can be seen that potential wells ($V < 0$) are found at those locations where the average magnetic shear $s$ is canceled by the distortions caused by the Pfirsch-Schl\"{u}ter current. Potential barriers ($V$ large and positive) are found in those regions where field line bending is most strongly inhibited due to increased local shear.

\subsection{Branches of ideal MHD Alfv\'{e}n eigenmodes}
\label{sec:ideal_branches}

The potential $V$ measures how much energy is needed to bend magnetic field lines. Increasing field line bending energy makes $V$ larger, while increasing free expansion energy associated with the pressure gradient $\alpha$ makes $V$ smaller. In the potential wells where $V$ is negative, free expansion energy exceeds field line bending energy locally. This occurs in regions with unfavorable curvature with respect to the direction of the pressure gradient when the local magnetic shear, $|h'|$, is sufficiently small.

A shear Alfv\'{e}n wave travelling along a field line and encountering a barrier will be reflected, at least partially. Thus, wave energy tends to be trapped between the barriers and standing waves can form whenever the wave function satisfies the quantization condition imposed by the barriers. There are two types of discrete solutions: purely growing eigenmodes and damped oscillatory eigenmodes. The damping of the latter is due to the finite width and finite height of the potential barriers, so there is either tunneling or direct coupling to waves propagating outward (corresponding to the continuous spectrum). The purely growing solutions are those which can tap a sufficient amount of free expansion energy to overcompensate field line bending energy. The Schr\"{o}dinger potential $V$ consists of an infinite number of potential barriers and wells, so the $s$-$\alpha$ plane contains overlapping bands of discrete shear Alfv\'{e}n eigenmodes localized in one or several potential wells. In the following, we shall consider some concrete examples.

\begin{figure}[tbp]
\includegraphics[width=1.00\textwidth]
{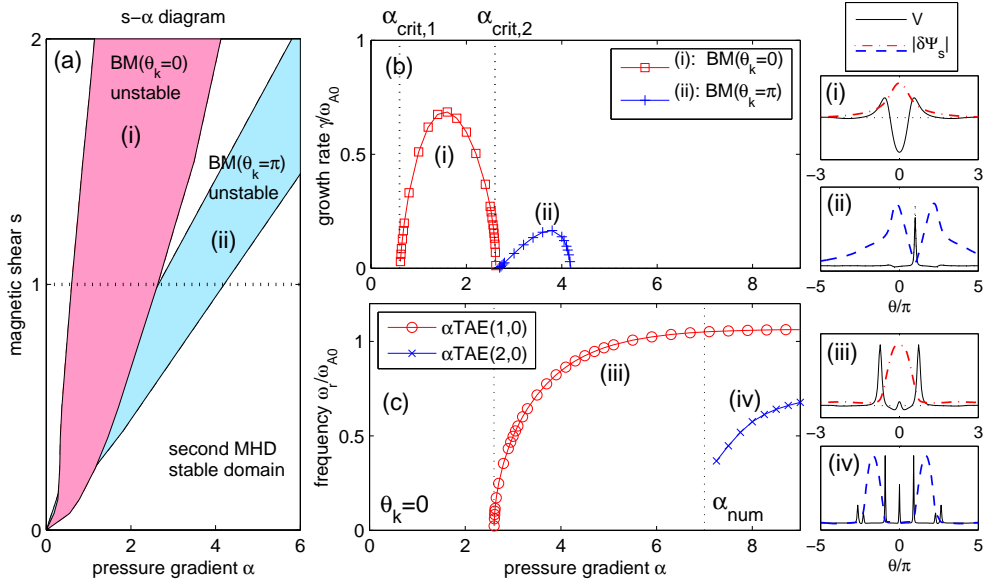}
\caption{Ideal MHD shear Alfv\'{e}n eigenmodes in the $s$-$\alpha$ plane. (a): Examples of ideal MHD ballooning unstable bands (BM=ballooning mode), for $\theta_k=0$ and $\theta_k=\pi$. (b): $\alpha$-dependence of the growth rates $\gamma$ of the BM bands $\theta_k=0$ (i) and $\theta_k=\pi$ (ii) for $s=1.0$. The vertical dotted lines indicate the first and second stability boundaries for $\theta_k=0$, which are labeled $\alpha_{\rm crit,1}$ and $\alpha_{\rm crit,2}$, respectively. (c): $\alpha$-dependence of the real frequencies $\omega_{\rm r}$ of $\alpha$TAE$(j,p)$ ground states ($p=0$) for $\theta_k=0$. $\alpha$TAEs localized in the central (iii) and in the second potential well (iv) are shown. $\alpha_{\rm num}$ is the numerical threshold above which band (iv) could be identified. On the right-hand side, diagrams (i)--(iv) show the typical structure of the Schr\"{o}dinger potential $V(\theta)$ and eigenmodes $|\delta\Psi_{\rm s}(\theta)|$ in covering space $-\infty < \theta < \infty$ for the bands (i)--(iv) shown in (b) and (c).}
\label{fig:bands}%
\end{figure}

Purely growing solutions are known as ideal MHD ballooning modes (BM). Two bands of purely growing ballooning modes, having different radial envelopes $\theta_k=0$ and $\theta_k=\pi$, are shown in figure~\ref{fig:bands}(a). Unstable bands with intermediate values $0 < |\theta_k| < \pi$ are located in the region spanned by the two bands shown \protect\cite{Chen87}. In figure~\ref{fig:bands}(b), the growth rates in these two bands are plotted as functions of $\alpha$ for $s=1.0$. The corresponding mode structures and potentials are plotted in figure \ref{fig:bands}(i) and (ii). The potential structure is symmetric around $\theta = \theta_k$ for integer values of $\theta_k/\pi$ and asymmetric otherwise.

For a given value of $s$, the ballooning branch appears at $\alpha_{\rm crit,1}(s)$ [$\approx 0.6$ for band (i) in figure \ref{fig:bands}(b)] and merges back into the accumulation point $\omega = 0$ at the second stability boundary, $\alpha_{\rm crit,2}(s)$ [$\approx 2.6$ for band (i) in figure \ref{fig:bands}(b)], as field line bending again balances free expansion energy. As $\alpha$ increases past this point, the purely growing ballooning branch is replaced by an oscillatory bound state localized in the same potential well, $-\pi \lesssim \theta \lesssim \pi$. This $\alpha$TAE branch is labeled (iii) in figure \ref{fig:bands}(c), where the eigenfrequency is plotted as a function of $\alpha$. As $\alpha$ increases further, another $\alpha$TAE localized in the second potential well, $\pi \lesssim |\theta| \lesssim 3\pi$, is found. In figure \ref{fig:bands}(c), this branch is shown for $\alpha > 7$ and labeled (iv). The typical mode structures are plotted in figure \ref{fig:bands}(iii) and (iv) along with the corresponding potential $V$.

The range of $\alpha$ values scanned exceeds by far the range of validity of the $s$-$\alpha$ model and probably goes far beyond what is realistically achievable in a tokamak-type plasma. However, this exaggeration is necessary to reveal the band structure of the $s$-$\alpha$ plane and is useful to study the properties of the shear Alfv\'{e}n eigenmodes which populate it.

\subsection{Characterization of $\alpha$-induced toroidal Alfv\'{e}n eigenmodes ($\alpha$TAE)}
\label{sec:ideal_atae}

In general, $\alpha$TAEs consist of a propagating component, which belongs to the Alfv\'{e}n continuum, and a standing wave component trapped between potential barriers induced by the pressure gradient measured by $\alpha$. $\alpha$TAEs can exist as bound or unbound states. Mathematically, the distinction is made on the basis of the location of turning points in the complex $\theta = \theta_{\rm r} + i\theta_{\rm i}$ plane of the Stokes diagram \cite{White} (cf.~figure 9 in \cite{Hu05}). For a bound state, the turning points associated with the trapping potential barriers lie on the $\theta_{\rm r}$ axis, whereas turning points of unbound states have $\theta_{\rm i} \neq 0$. In other words, for unbound states, the reflected component is coupled to the continuum directly (and, thus, suffers strong damping), whereas bound states transmit energy to the continuum only via barrier tunneling.

Since there is an infinite number of potential barriers, there exists a dense spectrum of $\alpha$TAEs. However, in practice, only a few modes are distinguishable from the continuum, and these are found only for sufficiently large values of $\alpha$; i.e., primarily in the second MHD stable domain. Following the notation introduced in \cite{Hu04}, $\alpha$TAE branches are denoted as $\alpha$TAE$(j,p)$. The label $j \geq 1$ identifies the potential well where the mode peaks, with potential barriers approximately located at $\theta \sim \pm \pi (2|j-1| - 1)$ and $\theta \sim \pm \pi (2j - 1)$. The label $p \geq 0$ counts the number of zeros the mode structure has in potential well $j$ and corresponds to the energy quantum number, with $p=0$ being the ground state. The eigenfrequency is denoted by $\omega_{\alpha{\rm TAE}}^{(j,p)}$. The two branches in figure~\ref{fig:bands}(c) are $\alpha$TAE$(1,0)$ [branch (iii)] and $\alpha$TAE$(2,0)$ [branch (iv)].

Generally speaking, reflections off $\alpha$-induced potential barriers occur for any $\alpha > 0$; first, they form unbound states when $\alpha$ is small; later, when $\alpha$ is sufficiently large, these turn into bound states. The $\alpha$TAE$(1,0)$ branch is a special case: it is replaced by the unstable ideal MHD ballooning branch in the region $\alpha_{\rm crit,1} \leq \alpha \leq \alpha_{\rm crit,2}$. For sufficiently large values of $s \gtrsim 0.5$, the $\alpha$TAE$(1,0)$ branch reappears at $\alpha = \alpha_{\rm crit,2}$ in the form of a strongly bound state.

One may imagine similar bands of ideal MHD ballooning modes in potential wells with $j>1$, but no such modes were found in the $s$-$\alpha$ model equilibrium. Thus, as $\alpha$ is increased, $\alpha$TAE branches with $j>1$ (as well as states with $j=1$ and $p>0$) smoothly transform from an unbound to a weakly bound and, eventually, strongly bound state. When solving equation (\ref{eq:model_saw_eq}) with the shooting method, $\alpha$ has to exceed a certain numerical threshold, $\alpha_{\rm num}$, beyond which an $\alpha$TAE is sufficiently deeply trapped to be clearly identifiable as a distinct discrete eigenmode. In the case of the $\alpha$TAE(2,0) branch in figure \ref{fig:bands}(c), we have $\alpha_{\rm num} \approx 7$.

As $\alpha$TAEs become strongly bound (``quasi-marginally stable'') with increasing $\alpha$, they attain high frequencies comparable to the Alfv\'{e}n frequency. The two examples (iii) and (iv) in figure \ref{fig:bands}(c) illustrate this fact. Naturally, modes with larger energy quantum number $p$ tend to reach higher frequencies due to stronger field line bending. In the $s$-$\alpha$ plane, branches with different $j$ and $p$ can become degenerate by crossing or merging (not shown). For instance, the even $(1,0)$ branch merges with the odd $(1,1)$ branch when $\alpha$ becomes sufficiently large, because a potential barrier appears at $\theta = 0$.

\subsection{Effect of the radial envelope}
\label{sec:ideal_radial}

Our use of the term ``bound state'' only refers to the trapping of waves along a flux tube. An inspection of the dependence of the eigenfrequencies on the minor radial coordinate, $r$, and the wave number of the radial envelope, $\theta_k$, is necessary to determine whether a given mode is a globally bound state or whether it propagates radially \cite{Hu05}. A necessary (not sufficient) condition for the existence of a globally bound state is $\partial\omega_{\rm r}/\partial\theta_k = 0$. This is only the case when the potential $V$ is symmetric around the location where the mode is locally trapped. For $\alpha$TAEs trapped in the central potential well ($j=1$) this is the case for $\theta_k=0$, and for the second well ($j=2$) the condition is satisfied for $\theta_k = \pi$. $\alpha$TAEs with $j>2$ cannot form globally bound states because $\partial\omega_{\rm r}/\partial\theta_k \neq 0$ for any $\theta_k$ (note that $\theta_k\rightarrow\theta_k\pm 2\pi$ implies $|j-1|\rightarrow |j-1 \mp 1|$). A higher excitation threshold due to radial propagation may be expected for these modes. Further global analysis goes beyond the scope of the present paper. In the remainder of this work, we let $\theta_k=0$ (flat radial envelope) in order to be able to make comparisons with earlier local gyrokinetic studies.

\section{Effect of FLR on the Schr\"{o}dinger potential and ${\bm \alpha}$TAE eigenfunctions}
\label{sec:struc}

In this section, we examine how FLR effects modify the effective Schr\"{o}dinger potential (section \ref{sec:flr_potential}) and $\alpha$TAE mode structures (section \ref{sec:flr_dpsi}). It is found that the structure of $V_{\rm eff}(\theta)$ is similar to the ideal MHD potential $V(\theta)$ when $\omega \approx \omega_{*p{\rm i}}$; otherwise, the potential structure is modified by the term $V_{{\rm FLR},\kappa}$ given in equation (\ref{eq:model_saw_vflr_wk}). In the eigenmode structure, no significant change is observed in the shape of the dominant bound state component. However, FLR terms may reduce the damping rate by moving additional turning points to the $\theta_{\rm r}$ axis of the Stokes diagram and strongly modify the waveform of the outward propagating component.

\begin{figure}[tbp]
\centering
\includegraphics[width=1.0\textwidth]
{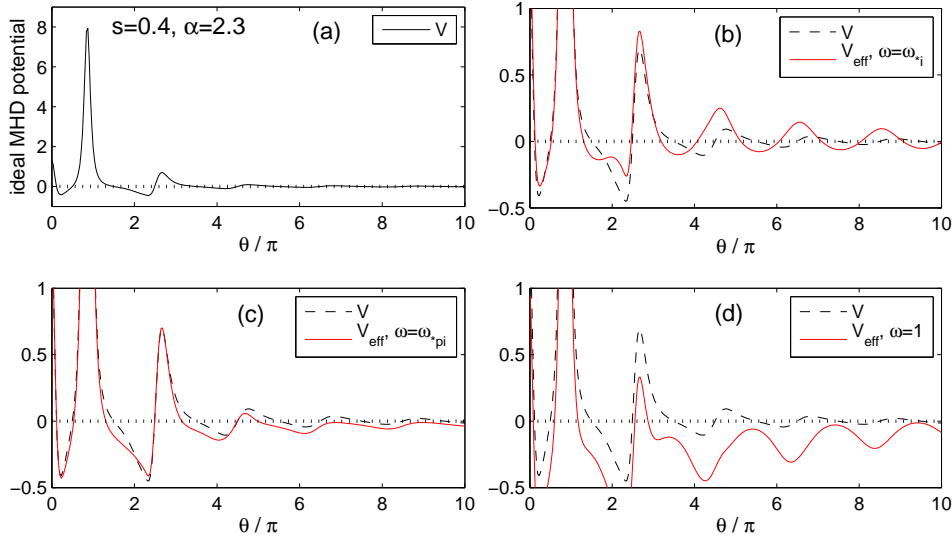}
\caption{Effective Schr\"{o}dinger potential $V_{\rm eff}$ for the case with $s=0.4$ in table \protect\ref{tab:parms} and $\alpha=2.3$. (a): Ideal MHD potential $V$. (b)--(d): $V_{\rm eff}$ (\full) in comparison with $V$ (\broken) for three values of $\omega$.}
\label{fig:flr_veff}%
\end{figure}

\subsection{Shape of the Schr\"{o}dinger potential}
\label{sec:flr_potential}

The shape of the ideal MHD potential $V(\theta|s,\alpha)$ is uniquely determined by the equilibrium geometry parametrized by $s$ and $\alpha$. When the FLR terms are included in equation (\ref{eq:model_saw_eq}), the effective potential acquires dependencies on many other parameters: $V_{\rm eff}(\theta|\omega,s,\alpha,\eta_{\rm i},\krhoi,\eps_n,q,\vti)$ [cf.~equation (\ref{eq:model_veff})]. In particular, note that the dependence on $\omega$ and $\krhoiO$ implies that modes with different frequencies and wavelengths observe different potential structures. This is illustrated in figure \ref{fig:flr_veff}, where the effective potential $V_{\rm eff}$ is shown for several values of $\omega$ and compared with the ideal MHD potential $V$. Despite this complexity, we can make several general remarks:
\begin{itemize}
\item  FLR terms may modify the shape of the bottom of a potential well (modifying the ballooning stability limits) and the shape of minor potential barriers (modifying their wave trapping properties). In the absence of kinetic compression, the effect on ballooning modes is found to be a stabilizing one when $\eta_{\rm i} > 0$ \cite{Tang81}.

\item  The potential $V_{\rm eff}$ is closest to its ideal-MHD limit $V$ when $\omega \sim \omega_{*p{\rm i}}$, as can be seen in figure \ref{fig:flr_veff}(c). This is particularly true for $|\theta| \gg 1/|\krhoiO s|$, as can be seen from the form of $Y(\omega)$ in equation (\ref{eq:saw_eq_flr_bc_coeff}) [which orginates from $V_{{\rm FLR},\kappa}$ in equation (\ref{eq:model_saw_vflr_wk}).

\item  FLR terms add offsets to the eigenfrequency, such as the well-known diamagnetic frequency shift. As is shown in section \ref{sec:scans_wf_shift} below, the effective value of this offset can vary from one potential well to the next, so the frequency shifts depend on the mode structure (i.e., the location of the dominant bound state component).
\end{itemize}

\begin{figure}[tbp]
\centering
\includegraphics[width=1.0\textwidth]
{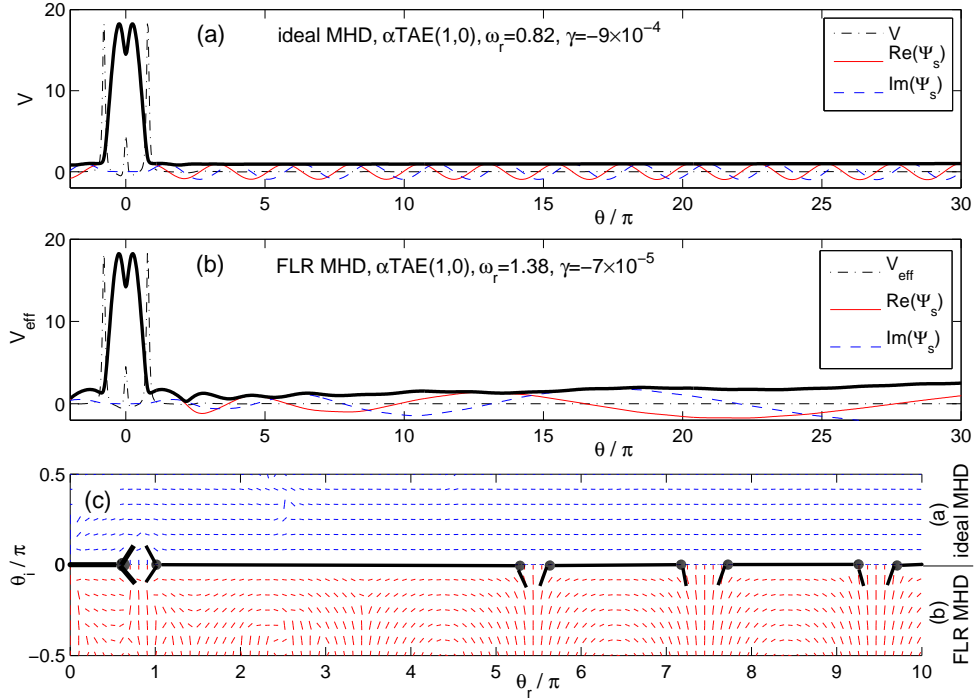}
\caption{Mode structure of a quasi-marginally stable $\alpha$TAE$(1,0)$ for the case $s=1.0$ (cf.~table \protect\ref{tab:parms}) with $\alpha=3.9$. Results are shown for (a) the ideal MHD limit [equation (\protect\ref{eq:saw_ideal})] and (b) for FLR MHD [equation (\protect\ref{eq:model_saw_eq})], where $\omega_{*p{\rm i}} = 0.88$. The respective effective Schr\"{o}dinger potential $V_{\rm eff}$ (\chain) is plotted along with the real (\full) and the imaginary component of $\delta\Psi_{\rm s}$ (\broken). Panel (c) shows the Stokes diagram with local anti-Stokes lines (\broken) for the ideal MHD (top half) and FLR MHD case (bottom half). In addition, global anti-Stokes lines which lie on the $\theta_{\rm r}$ axis (bold \full) and the associated turning points (\fullcircle) are indicated. Anti-Stokes lines represent paths in the complex $\theta = \theta_{\rm r} + i\theta_{\rm i}$ plane along which $\int{\rm d}\theta\, Q^{1/2}$ is imaginary (here, $Q = - V_{\rm eff}$, $\omega = \omega_{\rm r}$); i.e., where the eikonal approximation yields purely oscillatory solutions \protect\cite{White}.}
\label{fig:mstruc_10}%
\end{figure}

\begin{figure}[tbp]
\centering
\includegraphics[width=1.0\textwidth]
{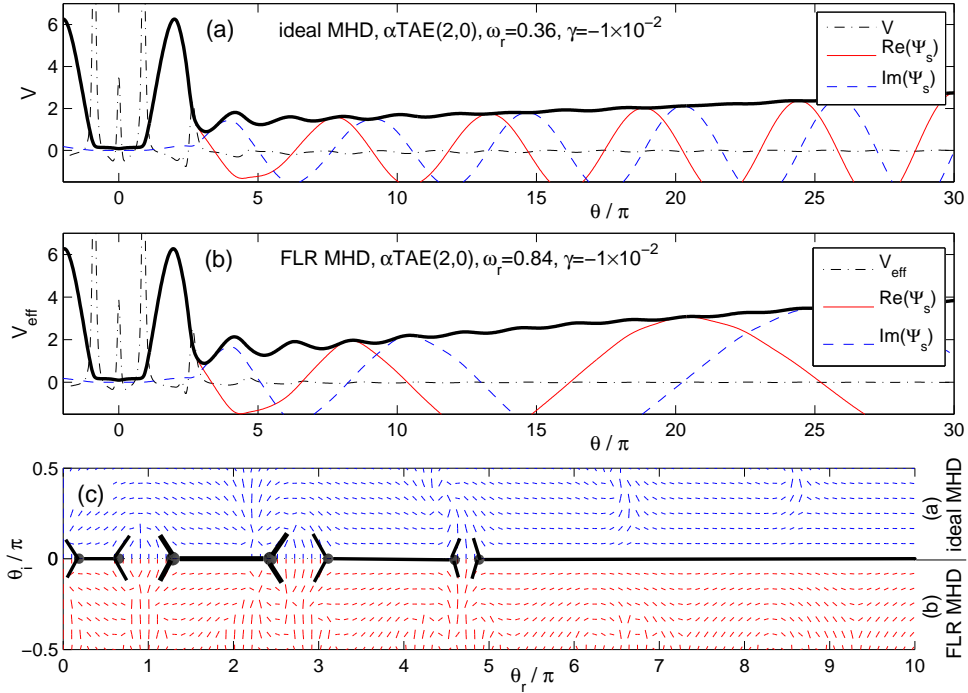}
\caption{Mode structure of a bound $\alpha$TAE$(2,0)$ for the case $s=0.4$ (cf.~table \protect\ref{tab:parms}) with $\alpha=3.0$ and $\omega_{*p{\rm i}} = 0.72$. The panels are arranged as in figure \protect\ref{fig:mstruc_10}.}
\label{fig:mstruc_20}%
\end{figure}

\subsection{$\alpha$TAE eigenfunctions}
\label{sec:flr_dpsi}

We analyze the effect of FLR terms on the mode structure of $\alpha$TAEs using the examples in figures \ref{fig:mstruc_10} and \ref{fig:mstruc_20}, where an $\alpha$TAE(1,0) for higher shear ($s=1.0$) and an $\alpha$TAE(2,0) for lower shear ($s=0.4$) are shown, respectively. The figures show (a) the ideal MHD result, (b) the FLR MHD result, and (c) the Stokes diagrams for the solutions in (a) and (b). As the Stokes diagrams show, the eigenmodes in both cases are strictly bound states; in fact, the $\alpha$TAE(1,0) in figure \ref{fig:mstruc_10} is already in the strongly bound (quasi-marginally stable) regime. The effective potential, $V_{\rm eff}$, and eigenvalues, $\omega = \omega_{\rm r} + i\gamma$, are also shown in panels (a) and (b) of both figures \ref{fig:mstruc_10} and \ref{fig:mstruc_20}.

No significant change in the structure of the dominant bound state components [identified by the labels $(j,p)$] is observed. Hence, we may assume that the increase in the eigenfrequency $\omega_{\rm r}$ in figures \ref{fig:mstruc_10} and \ref{fig:mstruc_20} is not due to a change in field line bending, but only an FLR-induced offset, which is discussed in section \ref{sec:scan} below.

Energy which tunnels through the potential barriers is carried away by the outgoing continuum wave which matches the bound state's frequency (resonant absorption). Most of the difference in the mode structures shown in figures \ref{fig:mstruc_10} and \ref{fig:mstruc_20} lies in the waveform of this propagating component:
\begin{itemize}
\item  Panel (a): Ideal MHD continuum waves are harmonic wave solutions obtained in the large-$|s\theta|$ limit of the ideal MHD equation (\ref{eq:saw_ideal}). The finite damping rate of an $\alpha$TAE is reflected in an exponentially growing amplitude of the form $\exp(-C \gamma|\theta|)$, where $C$ is a real positive number.

\item  Panel (b): The FLR continuum waves are governed by equation (\ref{eq:saw_eq_flr_bc}) obtained for $|\theta| \gg 1/|\krhoiO s|$ of the FLR MHD model (\ref{eq:model_saw_eq}). Distinctive features are an algebraically increasing amplitude (roughly $\propto |\theta|^{1/2}$) and an increasing phase velocity and wavelength (both $\propto |\theta|$) corresponding to the decay of the ion polarization current. A more detailed discussion is given in the Appendix.
\end{itemize}

\noindent These differences have to be taken into account when imposing boundary conditions in numerical codes used to study weakly bound $\alpha$TAEs such as those in figure \ref{fig:mstruc_20} (more about this in the Appendix). This is necessary to accurately compute the continuum damping, which may contribute to the instability threshold in the presence of wave-particle interactions, as is discussed in the companion paper \cite{Bierwage10b}. The eigenfrequencies are almost insensitive to the treatment of the outgoing continuum wave, because they are determined between the turning points associated with the primary potential well, $j$, where the main component of the $\alpha$TAE$(j,p)$ is trapped.

The bold solid lines in panels (a) and (b) of figures \ref{fig:mstruc_10} and \ref{fig:mstruc_20} represent the envelope of the mode structure. The modulation of this envelope indicates additional trapping of wave energy in secondary potential wells with $j+1$, $j+2$, etc. In the case with higher shear, shown in figure \ref{fig:mstruc_10}, we see that FLR terms significantly enhance the amount of trapping in secondary wells. The Stokes diagram in figure \ref{fig:mstruc_10}(c) reveals that, in the FLR MHD case, additional turning points are present on the $\theta_{\rm r}$ axis. This is consistent with the fact that the damping rate drops by one order of magnitude: from $-\gamma \approx 10^{-4}$ in ideal MHD [figure \ref{fig:mstruc_10}(a)] to $-\gamma \approx 10^{-5}$ in FLR MHD [figure \ref{fig:mstruc_10}(b)].

In contrast, in the case with lower shear, shown in figure \ref{fig:mstruc_20}, there is no significant change in the amount of trapping in secondary wells; the envelope $|\Psi_{\rm s}|$, the Stokes diagram, and the damping rates are similar in ideal MHD (a) and FLR MHD (b). According to the Stokes diagram in figure \ref{fig:mstruc_20}(c), both the ideal MHD and the FLR MHD eigenmode may be viewed as a combination of a primary (2,0) and a secondary (3,0) bound state component.

\section{Effect of FLR on ${\bm \alpha}$TAE eigenfrequencies}
\label{sec:scan}

In this section, we inspect how FLR effects modify the eigenfrequencies of $\alpha$TAEs. Eventually, the eigenfrequencies obtained here with the FLR MHD model (\ref{eq:model_saw_eq}) will allow us to identify Alfv\'{e}nic ITG instabilities seen in gyrokinetic simulation of the second ballooning stable domain \cite{Hirose94} as $\alpha$TAEs, which is done in a companion paper \cite{Bierwage10b}. With this motivation in mind, we study parameter scans with respect to $\alpha$, $\eta_{\rm i}$ and $\krhoiO$ in sections \ref{sec:scan_a_branches} and \ref{sec:scans_etai_k}. In section \ref{sec:scans_wf_shift} it is shown that, when $\eta_{\rm i} > 0$, the magnitude of the diamagnetic frequency shift depends on the mode structure; i.e., the location of the dominant bound state component.

\begin{figure}[tbp]
\includegraphics[width=1.00\textwidth]
{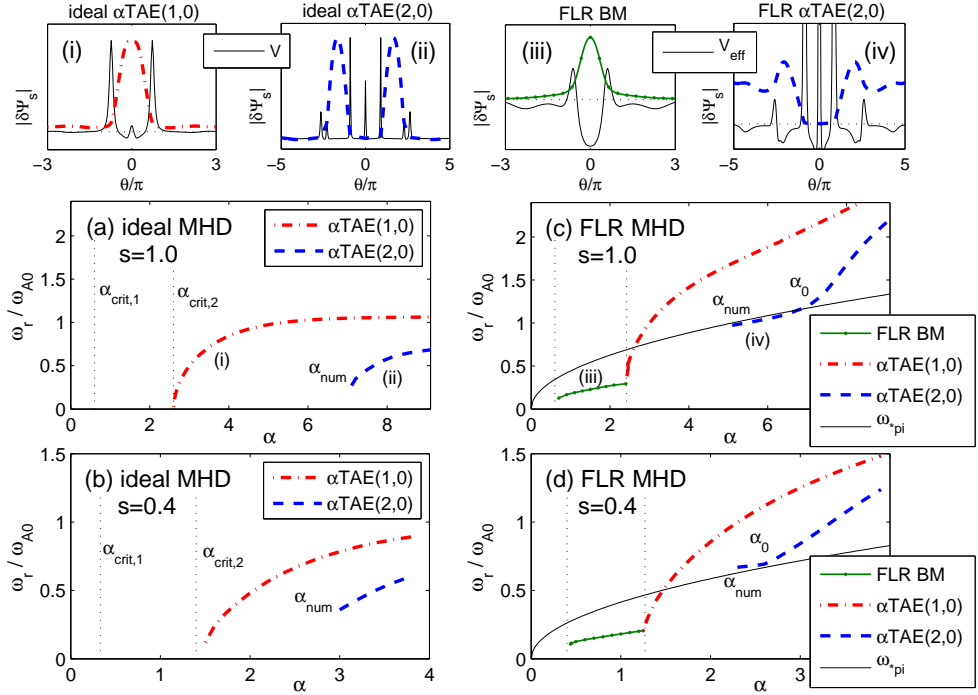}
\caption{$\alpha$-dependence of $\alpha$TAE eigenfrequencies in (a,b) ideal MHD and (c,d) FLR MHD for two values of magnetic shear: (a,c) $s=0.4$ and (b,d) $s=1.0$ (cf.~table \protect\ref{tab:parms}). In addition to the $\alpha$TAE branches (1,0) (\chain) and (2,0) (\broken), the frequency of the ideally unstable FLR ballooning mode (FLR BM) is shown in (c) and (d) (\fullcircle). For comparison, the diamagnetic frequency $\omega_{*p{\rm i}}$ is plotted in (c) and (d) (\full). The ballooning stability boundaries (\dotted) are labeled $\alpha_{\rm crit,1}$ and $\alpha_{\rm crit,2}$, the numerical threshold for identifying an $\alpha$TAE branch is labeled $\alpha_{\rm num}$, and the label $\alpha_0$ indicates the point at which the curve $\omega_{\rm r}(\alpha)$ exhibits a kink. Mode structure examples are shown in panels (i)--(iv).}
\label{fig:scan-a_branches}%
\end{figure}

\subsection{$\alpha$-dependence of the eigenfrequency}
\label{sec:scan_a_branches}

Figure \ref{fig:scan-a_branches} shows the frequencies of two branches, $\alpha$TAE$(1,0)$ (\chain) and $\alpha$TAE$(2,0)$ (\broken), as functions of the normalized pressure gradient $\alpha$. Ideal MHD results are shown in figure \ref{fig:scan-a_branches}(a) and (b) for $s=1.0$ and $s=0.4$, respectively. As discussed in section \ref{sec:ideal}, our method does not allow to follow the $\alpha$TAE(2,0) branch below a certain numerical threshold, $\alpha_{\rm num}$. Typical mode structures $|\delta\Psi_{\rm s}|$ found on the two branches are shown in panels (i) and (ii).

Corresponding results obtained with the FLR MHD model (\ref{eq:model_saw_eq}) are shown in figure \ref{fig:scan-a_branches}(c) and (d) where, in addition to the $\alpha$TAE branches, the frequency of the FLR-modified ballooning mode (FLR BM) is plotted as well (\fullcircle). For comparison, the diamagnetic frequency $\omega_{*p{\rm i}}$ is also shown (\full). Typical mode structures $|\delta\Psi_{\rm s}|$ found on the BM and $\alpha$TAE$(2,0)$ branch are shown in panels (iii) and (iv). Note that the numerical thresholds $\alpha_{\rm num}(s)$ are reduced in the FLR MHD case, so we are able to identify distinct $\alpha$TAE(2,0) bound states at lower values of $\alpha$ than in the ideal MHD limit. All $\alpha$TAE solutions shown in figure \ref{fig:scan-a_branches} are strictly bound states; with turning points on the $\theta_{\rm r}$ axis of the Stokes diagram (not shown).

\begin{figure}[tbp]
\includegraphics[width=1.00\textwidth]
{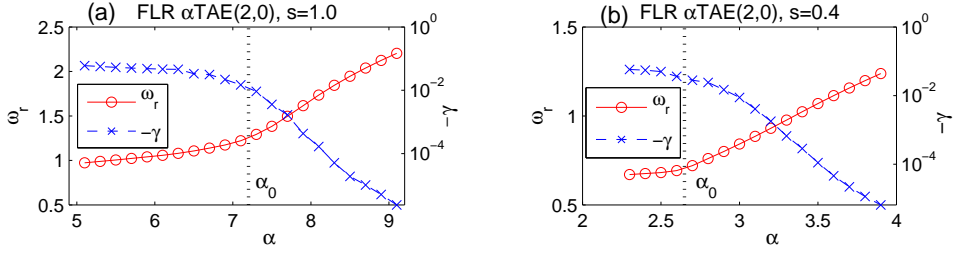}
\caption{FLR $\alpha$TAE(2,0) eigenfrequencies, $\omega_{\rm r}$ (\opencircle), and damping rates, $-\gamma$ ($\times$), as functions of $\alpha$ for (a) $s=1.0$ and (b) $s=0.4$ [cf.~figure \protect\ref{fig:scan-a_branches}(c) and (d)]. The location where the the slope of the curve $\omega_{\rm r}(\alpha)$ changes is labeled $\alpha_0$ (\dotted).}
\label{fig:alpha_kink}%
\end{figure}

When comparing the FLR MHD case [panels (c) and (d)] with the ideal MHD limit [panels (a) and (b)] in figure \ref{fig:scan-a_branches}, several interesting features can be observed:
\begin{itemize}
\item  As expected, the ballooning unstable domain shrinks and the eigenfrequencies undergo an up-shift. The frequency shift is primarily due to the diamagnetic frequency $\omega_{*p{\rm i}}$, with a significant contribution from $V_{{\rm FLR},\tau}$ due to finite $\delta E_\parallel$ (finite $\tauei^T$). When trying to quantify the frequency shift, we find that its value depends on the mode structure. This phenomenon is discussed in section \ref{sec:scans_wf_shift} below.

\item  Near the second FLR MHD ballooning stability boundary, the frequency of the FLR $\alpha$TAE(1,0) branch drops well below $\omega_{*p{\rm i}}$, which is known to be the continuum accumulation point of the diamagnetic gap in lowest-order FLR theories \cite{Tang81}. This observation is also explained in section \ref{sec:scans_wf_shift} below.

\item  In the FLR MHD case, the curves $\omega_{\rm r}(\alpha)$ for the $\alpha$TAE$(2,0)$ branch exhibit a kink at a location labeled $\alpha = \alpha_0$. For $\alpha < \alpha_0$, the frequency approximately scales like $\omega_{\rm r} \propto \alpha^{1/2}$; whereas a steep rise in the frequency occurs for $\alpha \gtrsim \alpha_0$. A comparison with the behavior of the damping rate $-\gamma(\alpha)$ in figure \ref{fig:alpha_kink} shows that, at least in the case with higher shear, $\alpha_0$ coincides with the threshold beyond which the bound state may be regarded as deeply trapped (``quasi-marginally stable'').
\end{itemize}

\begin{figure}[tbp]
\includegraphics[width=1.0\textwidth]
{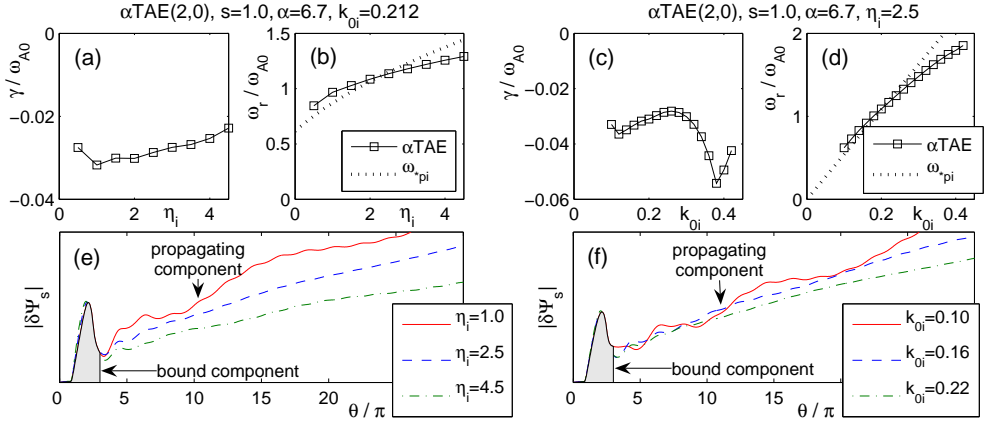}
\caption{Eigenvalues and eigenfunctions of an $\alpha$TAE(2,0) as functions of $\eta_{\rm i}$ (left panels) and $\krhoiO$ (right panels) for the case with $s=1.0$ (cf.~table \protect\ref{tab:parms}). The (negative) growth rates are shown in (a,c), the frequencies in (b,d), and mode structures in (e,f). For comparison, the diamagnetic frequency is plotted in (b) and (c) (\dotted). The shaded areas in (e) and (f) highlight the bound state component.}
\label{fig:scan-etai-k}%
\end{figure}

\subsection{$\eta_{\rm i}$- and $\krhoiO$-dependence of the eigenfrequency}
\label{sec:scans_etai_k}

Using the data point $\alpha = 6.7$ on the FLR $\alpha$TAE(2,0) branch in figure \ref{fig:scan-a_branches}(c) as a starting point, we vary the parameters $\eta_{\rm i}$ and $\krhoiO$. The results are shown in figure \ref{fig:scan-etai-k}.

In the entire parameter range scanned in figure \ref{fig:scan-etai-k}(a)--(d), the modes are strictly bound states and are subject to damping due to barrier tunneling; with $-\gamma/\omega_{\rm r} \sim 10^{-2}$. A sudden increase in the damping rate, such as near $\krhoiO \sim 0.4$ in figure \ref{fig:scan-etai-k}(c), typically indicates a degeneracy with or a switching to an $\alpha$TAE branch in a neighboring potential well, which is not necessarily evident in the eigenfrequency traces. In the case shown in figure \ref{fig:scan-etai-k}(c), an analysis of the Stokes diagram (not shown) reveals that the dip in $\gamma(\krhoiO)$ near $\krhoiO \sim 0.4$ is correlated with the merging of a turning point with the $\theta_{\rm r}$ axis near $|\theta| \sim 5\pi$; i.e., the formation of an additional bound state component $(j,p)=(3,0)$.

In figure \ref{fig:scan-etai-k}(b) and (d), the frequencies are found to be near $\omega_{*p{\rm i}}$ in most of the parameter range scanned. Although $\omega_{\rm r} \approx \omega_{*p{\rm i}}$ has special implications for the shape of the effective potential [cf.~section \ref{sec:flr_potential}], we have not yet found any conclusive evidence showing that this is indeed a preferred eigenfrequency for $\alpha$TAE in the regime $\alpha < \alpha_0$.

Note that the range of values of $\eta_{\rm i}$ and $\krhoiO$ examined in figure \ref{fig:scan-etai-k} is relevant for Alfv\'{e}nic ITG instabilities. Thus, the low damping rates and frequencies near the ion diamagnetic frequency suggest that these modes may be easily excited in the presence of ITG and kinetic compression, provided that $\omega_{\rm r}$ is also comparable to characteristic frequencies of particle motion. This is confirmed by gyrokinetic simulations \cite{Bierwage10b, Hirose94}.

\begin{figure}[tbp]
\centering
\includegraphics[width=1.0\textwidth]
{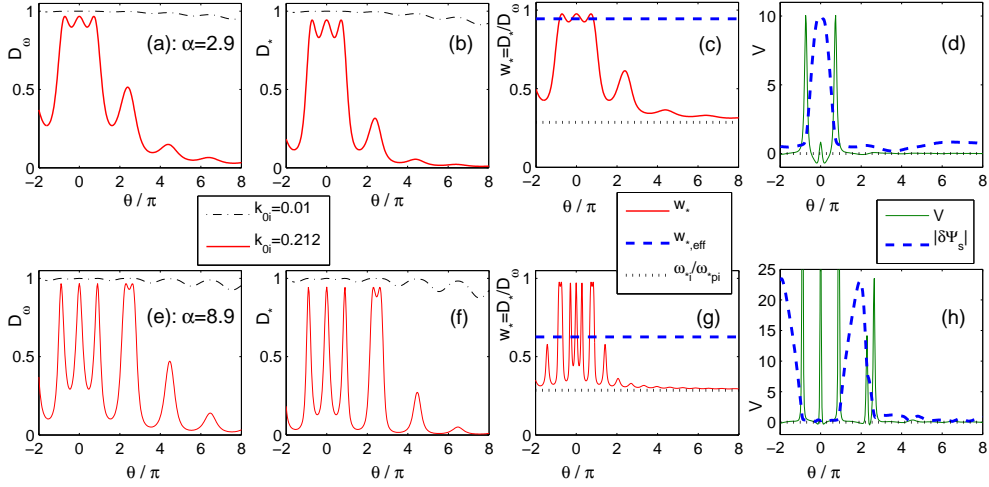}
\caption{Mode-structure-dependence of the diamagnetic frequency shift. The examples are based on the case $s=1.0$ in table \protect\ref{tab:parms} in the limit $\tauei^T=0$. In panels (a)--(d), results are shown for an $\alpha$TAE(1,0) at $\alpha=2.9$. In panels (e)--(h), results are shown for an $\alpha$TAE(2,0) at $\alpha=8.9$. Panels (a) and (e) show the spatial dependence of the coefficient $D_\omega=(1-\Gamma_0)/b_{\rm 0i}$, and (b) and (f) that of the coefficient $D_*=(1-\Gamma_0\Upsilon_1)/[b_{\rm 0i}(1+\eta_{\rm i})]$, for $\krhoiO = 0.01$ (\chain) and $\krhoiO = 0.212$ (\full). For $\krhoiO=0.212$ [where $\omega_{*p{\rm i}} = 0.45\sqrt{\alpha}$], (c) and (g) show the ratio $w_*=D_*/D_\omega$ (\full), its asymptotic limit for $|\theta| \rightarrow \infty$, which equals $1/(1 + \eta_{\rm i})$ (\dotted), and the normalized effective diamagnetic frequency $w_{*,{\rm eff}} = \overline{\omega_{*p{\rm i}}}/\omega_{*p{\rm i}}$ (\broken) as seen by the respective eigenfunction. The eigenfunctions $\delta\Psi_{\rm s}$ (\broken) are plotted in (d) and (h) together with the corresponding ideal MHD potential $V$ (\full).}
\label{fig:flr_inertia_coeff}%
\end{figure}

\subsection{Mode-structure-dependence of the diamagnetic frequency shift}
\label{sec:scans_wf_shift}

Low-order FLR models based on a small-$b_{\rm i}$ expansion, such as equation (\ref{eq:saw_flr_reduced}) in the Appendix, predict that the diamagnetic frequency causes an up-shift of the entire SAW spectrum and a gap in the continuous spectrum. The frequency shift depends only on $\omega_{*p{\rm i}}$. When FLR effects are fully retained, we find that the frequency shifts also depend on the mode structure, because the terms responsible for the shifts are functions of $\theta$, with a tendency to decay with increasing $|\theta|$. Typically, for modes localized in potential wells with $j > 1$, the full FLR model yields eigenfrequencies which are significantly lower than those obtained from low-order FLR models. Since the wave-particle resonance condition depends on the frequency, this effect is relevant for the resonant excitation of $\alpha$TAEs \cite{Bierwage10b}. In this section, mode-structure-dependence of the diamagnetic frequency shift in the FLR MHD model (\ref{eq:model_saw_eq}) is demonstrated through comparison between an $\alpha$TAE$(1,0)$ and an $\alpha$TAE$(2,0)$.

For simplicity, let us consider the case with $\delta E_\parallel = 0$. In this limit, which is realized by letting $\tauei^T=0$, the inertia term in equation (\ref{eq:model_maxw_vort1}) is
\begin{equation}
-V_{{\rm m},\omega}\delta\Psi_{\rm s} = \left(D_\omega \omega^2 - D_*\omega\omega_{*p{\rm i}}\right)\delta\Psi_{\rm s}.
\label{eq:flr_inertia1}
\end{equation}

\noindent After multiplication by the complex conjugate $\delta\Psi_{\rm s}^*$ and integration over $\theta$, we can define an effective diamagnetic frequency $\overline{\omega_{*p{\rm i}}}$ as
\begin{equation}
\omega(\omega - \overline{\omega_{*p{\rm i}}}) = \omega\left(\omega - \omega_{*p{\rm i}} \left<D_*\right>_\Psi/\left<D_\omega\right>_\Psi\right);
\label{eq:flr_inertia2}
\end{equation}

\noindent where the coefficients $D_\omega$ and $D_*$ are [cf.~equation (\ref{eq:model_saw_vflr_w})]
\begin{equation}
D_\omega = (1 - \Gamma_0)/b_{\rm i}, \quad
D_* = (1 - \Gamma_0\Upsilon_1)/[(1 + \eta_{\rm i})b_{\rm i}],
\label{eq:flr_inertia_coeff}
\end{equation}

\noindent and the brackets represent a weighted average,
\begin{equation}
\left<D_{*,\omega}\right>_\Psi = \left(\int_{\theta_1}^{\theta_2}{\rm d}\theta\, D_{*,\omega} |\delta\Psi_{\rm s}|^2\right)\left/\left(\int_{\theta_1}^{\theta_2}{\rm d}\theta\, |\delta\Psi_{\rm s}|^2/f\right) \right..
\label{eq:flr_inertia_coeff_eff}
\end{equation}

\noindent The limits of the integration interval, $[\theta_1,\theta_2]$, correspond to the turning points.

Two examples are presented in figure~\ref{fig:flr_inertia_coeff}. Based on the case with $s=1.0$ in table \ref{tab:parms}, results are shown for $\alpha=2.9$ (a)--(d) and $\alpha=8.9$ (e)--(h), where deeply trapped $\alpha$TAE(1,0) and $\alpha$TAE(2,0) are considered, respectively.

In figure~\ref{fig:flr_inertia_coeff}(a) and (b), $D_\omega$ and $D_*$ are plotted for $\krhoiO=0.01$ and $\krhoiO=0.212$, which clearly shows that FLR effects become important already at $|\theta|/\pi \sim \O(1)$. For $\krhoiO=0.212$, the ratio $w_* \equiv D_*/D_\omega$ is plotted in figure~\ref{fig:flr_inertia_coeff}(c) along with the normalized effective diamagnetic frequency, $w_{*,{\rm eff}} = \overline{\omega_{*p{\rm i}}}/\omega_{*p{\rm i}}$, as seen by the $\alpha$TAE(1,0) shown in figure~\ref{fig:flr_inertia_coeff}(d). The results for the $\alpha$TAE(2,0) in figure~\ref{fig:flr_inertia_coeff}(e)--(h) are arranged in the same manner.

It is found that, for the $\alpha$TAE(1,0) in figure~\ref{fig:flr_inertia_coeff}(d), the effective diamagnetic frequency amounts to 95\% of $\omega_{*p{\rm i}} = 0.76$, so there is no significant change. In contrast, for the $\alpha$TAE(2,0) in figure~\ref{fig:flr_inertia_coeff}(h), the effective diamagnetic frequency is only $63\%$ of $\omega_{*p{\rm i}} = 1.33$, so the associated frequency up-shift is correspondingly smaller.

The effect is also visible in figures \ref{fig:scan-a_branches}(c) and (d): at the second ballooning stability boundary, the transition between the FLR BM branch and the FLR $\alpha$TAE(1,0) branch occurs at a frequency well below $\omega_{*p{\rm i}}$, close to $\omega_{*{\rm i}} = \omega_{*p{\rm i}}/(1+\eta_{\rm i})$. This may be readily understood by noting that, near marginal stability, the solutions acquire a two-scale structure, consisting of a localized peak [$|k_\parallel| \sim 1/(q R_0)$] and a long-wavelength component ($|k_\parallel| \rightarrow 0$) \cite{Zonca96}. The long-wavelength component dominates in the spatial average given by (\ref{eq:flr_inertia_coeff_eff}), so that $\overline{\omega_{*p{\rm i}}}/\omega_{*p{\rm i}} \rightarrow 1/(1+\eta_{\rm i})$, which is the lower bound for the effective diamagnetic frequency. Obviously, this effect is only present for $\eta_{\rm i} > 0$, since $\Upsilon_1(\eta_{\rm i}=0)=1$ in equation (\ref{eq:flr_inertia_coeff}).

\section{Conclusion}
\label{sec:conclude}

The properties of $\alpha$-induced toroidal Alfv\'{e}n eigenmodes ($\alpha$TAE) have been investigated in detail using a fluid model including full finite-Larmor-radius (FLR) effects for thermal ions. The resulting changes in the effective Schr\"{o}dinger potential, the eigenmodes and eigenvalues were described and explained. By isolating the physics of the fluid limit, the results reported in the present paper provide the foundation for understanding how non-resonant and resonant effects of kinetic ion compression further modify the eigenvalues and eigenfunctions of $\alpha$TAEs.

$\alpha$TAEs become quasi-marginally stable and acquire high frequencies $\omega \gg \omega_{*p{\rm i}}$ when the normalized pressure gradient exceeds a branch-specific threshold, $\alpha > \alpha_0(j,p)$. In this regime, one may expect interactions with a population of energetic particles produced by auxiliary heating or nuclear fusion. Below this threshold, $\alpha < \alpha_0(j,p)$, continuum damping is stronger but the eigenfrequency tends to be near the diamagnetic frequency $\omega_{*p{\rm i}}$ and scales like $\alpha^{1/2}$. In this regime, $\alpha$TAEs may be expected to be easily excited via interactions with thermal ions in the presence of an ion temperature gradient (ITG). Indeed, as is shown in detail in the companion paper \cite{Bierwage10b}, FLR-modified $\alpha$TAEs are the normal modes which constitute the Alfv\'{e}nic ITG instabilities first reported by Hirose \textit{et al.} \cite{Hirose94} in the second ballooning stable domain.

\ack

One of the authors (A.B.) would like to thank Shuanghui Hu (Guizhou University) for helpful advice during the early stages of the project. He also thanks Masaru Furukawa (Tokyo University) and Zhihong Lin (UCI) for stimulating discussions. This research is supported by U.S.~DoE Grant DE-AC02-CH0-3073, NSF Grant ATM-0335279, and SciDAC GSEP.

\appendix
\section*{Appendix: Outward propagating wave in FLR MHD}
\setcounter{section}{1}
\label{sec:appendix_flr_continuum}

This appendix summarizes peculiarities of the FLR MHD model which were encountered during our study. This model-specific information is included as a reference for researchers trying to reproduce our results and understand them in detail.

\subsection*{Solution for the FLR continuum}

Equation (\ref{eq:saw_eq_flr_bc}) may be solved approximately with the \textit{ansatz} $\delta\Psi_{\rm s} = \overline{\psi} + \widetilde{\psi}\sin\theta$, where $\overline{\psi}$ is a long-wavelength component associated with the $X/\theta^2$ term and $\widetilde{\psi}$ is a short-wavelength component associated with the $Y\sin\theta/\theta$ term. We assume the following formal ordering with respect to the small parameter $\delta \sim 1/|\krhoiO s \theta| \ll 1$:
\begin{equation}
\fl \overline{\psi}'/\overline{\psi} \sim \O(\delta), \quad
\widetilde{\psi}'/\widetilde{\psi} \sim \O(\delta), \quad
\widetilde{\psi}/\overline{\psi} \sim \O(\delta), \quad
|\krhoiO s \theta| \sim \O(\delta^{-1});
\end{equation}

\noindent where the prime denotes a derivative with respect to $\theta$. Separating the equations with respect to the long and short wavelengths, we obtain, at lowest order,
\begin{equation}
0 = \overline{\psi}'' + \frac{X}{(\krhoiO s \theta)^2}\overline{\psi} + \frac{Y}{2\krhoiO s \theta}\widetilde{\psi}, \qquad
\widetilde{\psi} = \frac{Y}{\krhoiO s \theta} \overline{\psi}.
\label{eq:apdx_saw_eq}
\end{equation}

\noindent The solution for $\overline{\psi}$ has the form $|\krhoiO s \theta|^{(1 \pm iC)/2}$ with $C = [(4X + 2Y^2)/(\krhoiO s)^2 - 1]^{1/2}$, so that
\begin{equation}
\fl \delta\Psi_{\rm s} \propto |\krhoiO s \theta|^{\frac{1}{2} \pm i\frac{C}{2}} \left(1 + \frac{Y\sin\theta}{\krhoiO s \theta}\right), \quad
k_\parallel \approx \frac{1}{2\theta} \left[\pm C - i\left(1 + \frac{2Y\cos\theta}{\krhoiO s}\right)\right];
\label{eq:kpar}
\end{equation}

\noindent where $k_\parallel \equiv -i\delta\Psi_{\rm s}'/\delta\Psi_{\rm s}$ and, typically, $C^2 > 0$. Equation (\ref{eq:kpar}) describes the propagating component of an eigenmode which couples to the local shear Alfv\'{e}n continuum in FLR MHD, either through direct coupling or barrier tunneling.

In order to determine the physically relevant sign in the exponent, we compute the group velocity $v_{\rm g} = {\rm d}\omega/{\rm d}k_\parallel$. For simplicity, let us first consider a special type of solutions characterized by $\omega \approx \omega_{*p{\rm i}}$. In this case, $Y\approx 0$ and we have
\begin{equation}
{\rm Re}\left\{\frac{{\rm d}k_\parallel}{{\rm d}\omega_{\rm r}}\right\} \approx \pm\frac{1}{2\theta} \left[\frac{{\rm d}}{{\rm d}\omega_{\rm r}}{\rm Re}\{C\}\right]_{\omega = \omega_{*p{\rm i}}}, \nonumber
\end{equation}

\noindent so that
\begin{equation}
v_{\rm g} \approx \pm 2|\krhoiO s| \theta D;
\label{eq:vgroup}
\end{equation}

\noindent where $D > 0$ is a constant. The outgoing wave has $v_{\rm g} > 0$ for $\theta > 0$ (and vice versa), so that the physically relevant solution to be matched at the boundary is the one with the positive sign in the exponent as in equation (\ref{eq:match_flr}).

\begin{figure}[tbp]
\centering
\includegraphics[width=1.0\textwidth]
{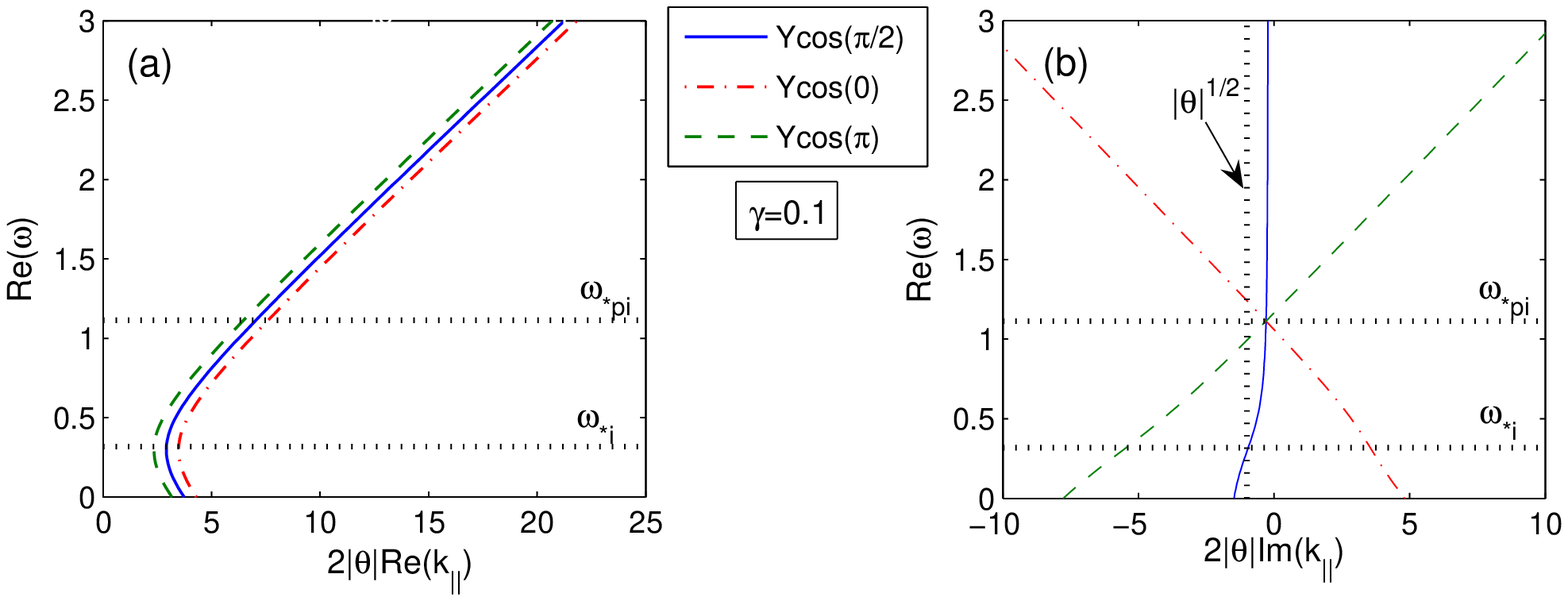}
\caption{(a) Dispersion relation and (b) modulation of the algebraically growing amplitude for outgoing waves at large $|\theta|$. Results are shown for values of $\theta$ where the short-wavelength contribution $\propto Y\cos\theta$ in equation (\protect\ref{eq:kpar}) vanishes (\full), is maximal (\chain) and minimal (\broken). The parameters for the case with $s=1.0$ in table \protect\ref{tab:parms} are used, and we set $\gamma=-0.1$, $\alpha=6.3$. For comparison, horizontal dotted lines in (a) and (b) indicate $\omega_{*p{\rm i}}$ and $\omega_{*{\rm i}}$, and the vertical dotted line in (b) marks the average algebraic growth associated with ${\rm Im}\{k_\parallel\}$ for the case $\gamma = 0$, which is $|\theta|^{1/2}$.}
\label{fig:dispersion}%
\end{figure}

The WKB dispersion relation $k_\parallel = k_\parallel(\theta|\omega_{\rm r})$ for any $\omega_{\rm r}$ is plotted in figure \ref{fig:dispersion}(a). The growth rate $\gamma$ enters mainly via the short-wavelength correction $\propto 2Y\cos\theta/(|\krhoiO s |\theta)$ in equation (\ref{eq:kpar}), the real part of which vanishes when $\omega_{\rm r} = \omega_{*p{\rm i}}$. Although its magnitude is usually small, this short-wavelength correction can still be important for matching at the boundary because it determines the local slope of the wave function (unless $\omega_{\rm r} \sim \omega_{*p{\rm i}} \gg |\gamma|$). The short-wavelength modulation of the algebraic growth associated with ${\rm Im}\{k_\parallel\}$ is shown in figure \ref{fig:dispersion}(b). For $\gamma = 0$, the average algebraic growth goes as $|\theta|^{1/2}$, as can also be seen from equation (\ref{eq:kpar}).

\subsection*{Properties of the FLR continuum wave}

The properties of a full FLR model at large $|\theta|$ were previously analyzed by Connor \textit{et al.} \cite{Connor83} in a simpler geometry (sheared slab), which exhibits similar qualitative features as the toroidal flux tube model used here. In particular, since the inertia term in equation (\ref{eq:saw_eq_flr_bc}) disappears in the limit $|\krhoiO s \theta| \gg 1$, the full FLR model is very different from widely used lowest-order FLR models which are based on the assumption $b_{\rm i} \ll 1$. As an example, consider the model used in the first studies of so-called ``kinetic ballooning modes'' (in this work, called FLR ballooning modes) with $\omega \sim \omega_{*p{\rm i}}$ \cite{Tang81, Cheng82, Tsai93} [normalized as in equation~(\ref{eq:norm_omgf_omgk})]:
\begin{equation}
0 = \delta\Psi_{\rm s}'' + \omega(\omega - \omega_{*p{\rm i}})\delta\Psi_{\rm s} - V\delta\Psi_{\rm s}.
\label{eq:saw_flr_reduced}
\end{equation}

\noindent In the limit $|s\theta| \gg 1$, where $V \propto 1/(s\theta)^2$ vanishes, equation (\ref{eq:saw_flr_reduced}) describes harmonic waves. In contrast, the solutions (\ref{eq:kpar}) have the following notable properties:
\begin{itemize}
\item  The parallel ``wavelength'' ($\sim 1/|k_\parallel|$), and both the phase and the group velocity are all proportional to $\theta$; whereas, in the lowest-order FLR model, they are constant.

\item  The wave amplitude exhibits algebraic growth proportional to $|\theta|^{1/2}$. The exponential increase/decrease in the magnitude of damped/growing solutions (due to ${\rm Im}\{k_\parallel\} \propto \gamma \lessgtr 0$) known from the lowest-order FLR model is also turned into algebraic increase/decrease in the full FLR model.
\end{itemize} 

\noindent These differences can be seen in figure \ref{fig:mstruc_10} which shows shooting code results for $\alpha$TAE trapped in the central potential in (a) the ideal MHD limit [equation (\ref{eq:saw_flr_reduced}) with $\omega_{*p{\rm i}} = 0$] and (b) the full FLR MHD model [equation (\ref{eq:model_saw_eq})]. Here, the bound state component is located in the region $|\theta| \lesssim \pi$ and the outgoing continuum wave in the region $|\theta| > \pi$.

\begin{figure}[tbp]
\centering
\includegraphics[width=1.0\textwidth]
{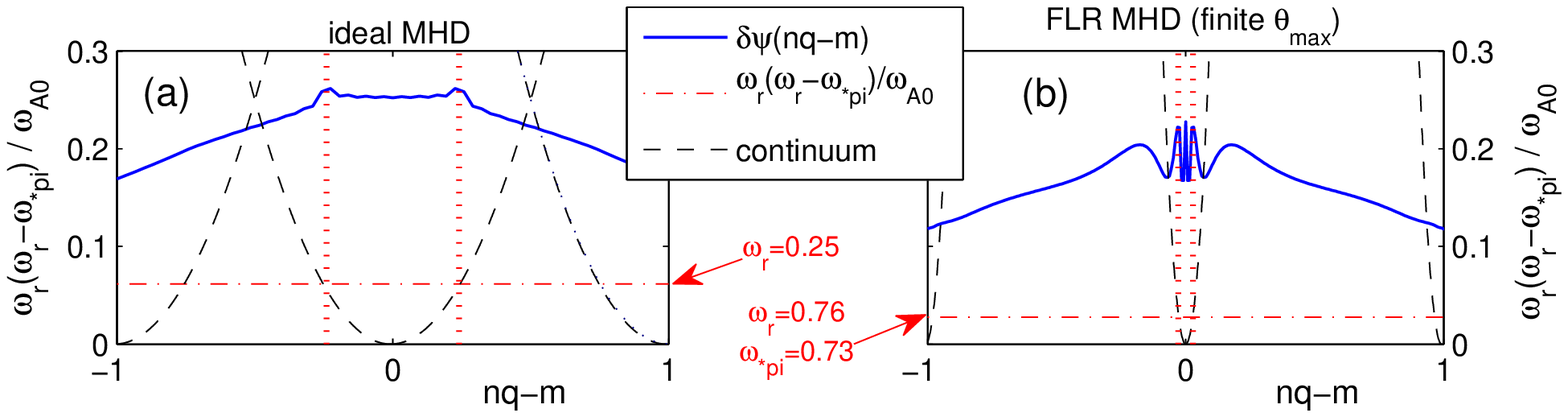}
\caption{Radial mode structure $\delta\psi(nq-m)$ (\full) of an $\alpha$TAE$(1,0)$ in the case with $s=1.0$ in table \protect\ref{tab:parms} for $\alpha=2.7$. The ideal MHD (a) and the FLR MHD solution (b) is shown. The mode structure exhibits singularities (\dotted) at the locations where the effective mode frequency $\sqrt{\omega_{\rm r}(\omega_{\rm r} - \omega_{*p{\rm i}})}$ (\chain) matches the frequency of the local continuum (\broken). In (b), the continuum frequency (\broken) diverges as $\theta_{\rm max} \rightarrow \infty$ (which translates to $t \rightarrow \infty$ due to finite propagation velocity), but it is plotted here for a finite domain size to fit the numerical result for $\delta\psi(nq-m)$.}
\label{fig:mstruc_nqm}%
\end{figure}

After Fourier transformation, the mode structure $\delta\psi(nq-m)$ exhibits singular spikes at radii satisfying $|n q(r) - m|^2 = \omega_{\rm r}(\omega_{\rm r} - \overline{\omega_{*p{\rm i}}}) > 0$, where $\overline{\omega_{*p{\rm i}}}$ is the effective diamagnetic frequency defined in equation (\ref{eq:flr_inertia2}). In FLR MHD, the Alfv\'{e}n velocity diverges as $|\theta| \rightarrow \infty$, because the ion polarization current vanishes (and there is no electron inertia or collisionality in our model to replace it), so the locations of the singularities move to $n q(r) - m = 0$. The effect is illustrated in figure \ref{fig:mstruc_nqm}. Note that, since the propagation velocity is finite, $|\theta| \rightarrow \infty$ translates to $t \rightarrow \infty$, which means that the continuum frequency sweeps up for ${\rm Re}(k_\parallel) \neq 0$.

\subsection*{Implications for the construction of quadratic forms}

The algebraic increase in the wave amplitude seen in figures~\ref{fig:mstruc_10}(b) and \ref{fig:mstruc_20} is consistent with the analytical result (\ref{eq:kpar}), so it is not a numerical artifact. However, this raises the question whether it is possible to construct physically meaningful quadratic forms for a model, which is not physically self-consistent for $|\theta| \gg 1/|\krhoiO s|$. In other words, the question is whether and how square-integrability is satisfied when these solutions are destabilized by kinetic effects, as is done in \cite{Bierwage10b}. As will be shown in this section, the answer lies in the fact that, unlike in the ideal-MHD limit given by equation (\ref{eq:saw_ideal}), the magnitude of $|\delta\Psi_{\rm s}|^2$ is \textit{not} an appropriate measure for the wave energy density.

Written in quadratic form, equation (\ref{eq:model_saw_eq}) becomes
\begin{equation}
i\Phi_{\rm b} = \delta\L;
\label{eq:model_saw_dw}
\end{equation}

\noindent where, for the simulation domain $\theta \in [-\theta_{\rm max}, \theta_{\rm max}]$,
\begin{eqnarray}
\Phi_{\rm b} &=& i\left[\delta\Psi_{\rm s}^*\delta\Psi_{\rm s}'\right]_{-\theta_{\rm max}}^{\theta_{\rm max}},
\\
\delta\L &=& -\int_{-\theta_{\rm max}}^{\theta_{\rm max}} {\rm d}\theta\, \left[ \left|\delta\Psi_{\rm s}'\right|^2 + V\left|\delta\Psi_{\rm s}\right|^2\right]
\label{eq:apdx_lagrangian}
\\
&& - \int_{-\theta_{\rm max}}^{\theta_{\rm max}} {\rm d}\theta\, \left(V_{{\rm m},\omega} + V_{{\rm m},\tau} + V_{\kappa,{\rm FLR}} \right)\left|\delta\Psi_{\rm s}\right|^2. \nonumber
\end{eqnarray}

\noindent Here, $\Phi_{\rm b}$ measures the energy flux through the boundary and $\delta\L$ is the Lagrangian (see equation (31) in \cite{Zonca93}). The structure of $V_{\rm eff}$ and $\delta\Psi_{\rm s}'$ indicates that $|\delta\psi|^2 = |\delta\Psi_{\rm s}|^2/f$ represents the wave energy density, even though $\delta\Psi_{\rm s}$ is the relevant wave function; i.e., a solution of the SAW Schr\"{o}dinger equation (\ref{eq:model_saw_eq}).

Our simulations using \textsc{awecs} \cite{Bierwage08} confirm that $|\delta\psi|^2$ decays like $1/|\theta|^x$, with $x \approx 1$ for marginally stable modes and $x>1$ for unstable modes. Thus, unstable modes are square-integrable and it is meaningful to use quadratic forms for theoretical arguments and for analysis of simulation results, which is done in a companion paper \cite{Bierwage10b}.

The above implies that the appropriate normalization constant for the quadratic forms is $\int{\rm d}\theta |\delta\Psi_{\rm s}|^2/f$ [as in equation (\ref{eq:flr_inertia_coeff_eff})]. Although the normalization constant is irrelevant for any single case, the choice is crucial for plotting meaningful results for parameter scans \cite{Bierwage10b}.

Note that the first line of equation (\ref{eq:apdx_lagrangian}) corresponds to $2\delta W_{\rm f}$ which measures the ideal MHD potential energy and the $\omega^2$ term on the second line measures the ideal MHD inertia. However, since the eigenvalue of an $\alpha$TAEs is determined in the region between the turning points of the potential barriers, there is no two-scale structure and no distinction can be made between an ``ideal region'' and an ``inertial layer'' (as used, e.g., in \cite{Zonca96}), so we prefer not to separate kinetic and potential energies and keep them unified in the Lagrangian $\delta\L$ \cite{Bierwage10b}.

\subsection*{Implications for numerical simulation of a linear initial-value problem with $\alpha$TAEs}

Numerical simulations are always done in a finite-size domain, so outward propagating waves must be absorbed by appropriate boundary conditions in order to avoid unphysical reflections. Here, we discuss two methods which may be used in linear gyrokinetic initial-value simulations of $\alpha$TAEs:
\begin{itemize}
\item  \textit{Matching:} In our shooting code, we successfully match the numerical solution to a known analytical form given by equation (\ref{eq:kpar}) at the boundaries $\theta = \pm\theta_{\rm max}$ of the computational domain. This is not as easily done in gyrokinetic initial value simulations, since the boundary condition given by equation (\ref{eq:kpar}) is a function of the eigenvalue $\omega$, which is not known \textit{a priori}. One may attempt to overcome this constraint by an iterative or adaptive approach which, however, may adversely affect the stability of numerical finite-difference and integration schemes, and appears to require additional artificial damping.

\item  \textit{Absorbing boundary:} A popular method used in initial value codes is the absorbing boundary condition, where one imposes an artificial damping rate in the outer reaches of the simulation domain. However, this is effective only for those components of the wave, which have wavelengths smaller than the damping region. For the dominant long-wavelength component of the FLR continuum waves described by equation (\ref{eq:kpar}), such an absorption layer acts like a reflecting wall.
\end{itemize}

These examples illustrate that it is difficult to treat FLR continuum waves accurately in initial-value codes. Fortunately, it is often possible to tolerate unphysical reflections off the boundary; in particular, when the following constraints are satisfied:
\begin{itemize}
\item  The mode is driven strongly unstable. By the time the continuum wave returns, its energy is exponentially small compared to the bound state.

\item  The mode is unstable and deeply trapped between large potential barriers, so the fraction of returning wave energy which penetrates the barriers is exponentially small.
\end{itemize}

\noindent Clearly, the most challenging case is the simulation of weakly bound states near marginal stability.

\setlength{\bibsep}{0.6pt}
\bibliographystyle{unsrt}
\bibliography{p08_flr_atae}

\end{document}